\documentclass[preprint]{aastex}
\usepackage{color}
\usepackage{graphics}
\usepackage{fancyheadings}
\usepackage{lscape}
\usepackage{longtable,lscape}
\usepackage{rotating}
\usepackage{epsfig}
\usepackage{subfigure}
\usepackage{multirow}
\usepackage{amsmath}
\usepackage{mathtools}

\begin{document}

\newcommand{\dg}{$^{\circ}$} 
\newcommand{\sex}{{\it SExtractor}}
\newcommand{\Msol}{M$_{\odot}$}
\newcommand{\col}{$\colon$}
\newcommand{\longdash}{---}
\newcommand{\gsim}{$\gtrsim$}
\newcommand{\nod}{\nodata}
\newcommand{\lsim}{$\lesssim$}
\newcommand{\txw}{\textwidth}
\newcommand{\snr}{signal-to-noise ratio}
\newcommand{\hypz}{{\tt HyperZ}}
\newcommand{\gfit}{{\tt GalFit}}
\newcommand{\Sersic}{S\'{e}rsic}
\newcommand{\rsrang}{$0.35$\lsim$z$\lsim$1.5$}
\newcommand{\nuvv}{(NUV--V)$_{r}$}
\newcommand{\fuvv}{(FUV--V)$_{r}$}
\newcommand{\nuvr}{(NUV--r$^{\prime})_{r}$}
\newcommand{\fuvr}{(FUV--r$^{\prime})_{r}$}
\newcommand{\gmr}{($g^{\prime}-r^{\prime}$)$_{r}$}

\title{\bf Early-type galaxies at intermediate redshift observed with
  HST WFC3\col perspectives on recent star-formation.}

\author{Michael J. Rutkowski\altaffilmark{1}, Hyunjin
  Jeong\altaffilmark{2}, Seth H. Cohen\altaffilmark{3}, Sugata Kaviraj\altaffilmark{5}, Russell
  E. Ryan, Jr.\altaffilmark{5}, Anton Koekemoer\altaffilmark{5}, Rogier A. Windhorst\altaffilmark{3},  Nimish
  P. Hathi\altaffilmark{6},  Michael A. Dopita\altaffilmark{7,8,9}, Sukyoung
  K. Yi\altaffilmark{2}}
\altaffiltext{1}{Minnesota Institute for Astrophysics, University of Minnesota, 116 Church St. SE, Minneapolis, MN 55455}
\altaffiltext{2}{Department of Astronomy, Yonsei University 134, Shinchon-dong, Sudaemun-gu, Seoul 120-179 Korea}
\altaffiltext{3}{School of Earth and Space Exploration, Arizona State University, Tempe, AZ 85287-1404, USA}
\altaffiltext{4}{University of Hertfordshire, Hatfield, AL10 9AB, UK}
\altaffiltext{5}{Space Telescope Science Institute, Baltimore, MD 21218, USA}
\altaffiltext{6}{Aix Marseille Universit\'{e}, CNRS, LAM (Laboratoire d'Astrophysique de Marseille) UMR 7326, 13388, Marseille, France}
\altaffiltext{7}{Research School of Physics and Astronomy, The Australian National University, Canberra ACT 2611, Australia}
\altaffiltext{8}{Astronomy Department, King Abdulaziz University, P.O. Box 80203, Jeddah, Saudi Arabia}
\altaffiltext{9}{Institute for Astronomy, University of Hawaii, Honolulu, HI 96822, USA}
\altaffiltext{10}{Department of Astronomy, Yonsei University 134, Shinchon-dong, Sudaemun-gu, Seoul 120-179 Korea}

\begin{abstract} 

We present an analysis of the stellar populations of 102
visually-selected early-type galaxies (ETGs) with spectroscopic
redshifts (\rsrang) from observations in the Early Release Science
program with the Wide Field Camera 3 (WFC3) on {\it Hubble Space
Telescope} (HST). We fit one- and two-component synthetic stellar models
to the ETGs UV-optical-near-IR spectral energy distributions and find a
large fraction ($\sim$40\%) are likely to have experienced a minor
($f_{YC}$\lsim10\% of stellar mass) burst of recent ($t_{YC}$\lsim 1
Gyr) star-formation. The measured ages and mass fraction of the young
stellar populations do not strongly trend with measurements of galaxy
morphology.  We note that massive (M$\,>10^{10.5}$\Msol) recently
star-forming ETGs appear to have larger sizes.  Furthermore, high-mass,
quiescent ETGs identified with likely companions populate a distinct
region in the size-mass parameter space, in comparison with the
distribution of massive ETGs with evidence of recent star-formation
(RSF).  We conclude that both mechanisms of the quenching of
star-formation in disk-like ETGs and (gas-rich, minor) merger activity
contribute to the formation of young stars and the size-mass evolution
of intermediate redshift ETGs. The number of ETGs for which we have both
HST WFC3 panchromatic (especially UV) imaging and
spectroscopically-confirmed redshifts is relatively small, therefore a
conclusion on the {\it relative} roles of both of these mechanisms
remains an open question. \end{abstract}

\section{Introduction}\label{sec:intro} 

Massive (M\gsim10$^{11}$\Msol) early-type galaxies (ETGs) dominate the
stellar mass and baryon budget in the low to intermediate-redshift
universe \citep[z\lsim1; see e.g.,][]{Fukujita98}.  However, their
assembly and evolution is not yet fully understood and is the subject of
active research. Generally, ETGs are observed to form a tight
``red-sequence'' in optical color-magnitude space and have long been
considered ``red and dead'' (deVaucouleurs 1961; Kennicutt et al.~1998).
 Furthermore, ETGs are tightly correlated on the fundamental plane
\citep{Bender92} and relatively enriched in $\alpha$-elements (e.g., Mg,
Ne; see Trager et al.~1998,\,2000; Thomas et al.~2005).  This evidence
has long been interpreted as support for a common formation scenario
\citep{Larson74,Pipino04,Chiosi02}, in which massive ETGs formed the
majority of their stellar mass at high redshift ($z\gg1$) in a
relatively short burst of star-formation
\citep{Matteucci94,Thomas99,Faber07}. This evolutionary history
contrasts with the more continuous star-formation histories observed for
later-type galaxies.

Recently, observations have revised the traditional picture of ETG
formation and evolution.   Rest-frame far- and near-ultraviolet (FUV and
NUV) observations---which are uniquely sensitive to recent
star-formation in quiescent ETGs \citep{Yi05}---have confirmed many
($\sim$30\%) low redshift ($z<0.3$) ETGs possess $\ll$ 5$-$10\%  of
their total stellar mass in young stellar populations (Ferreras \&
Silk~2000; Kaviraj et al.~2007b, 2008, 2011).  The presence of young
stars drives the optically-red, quiescent galaxies towards (perhaps
repeatedly) the ``green valley''~\citep{Wyder07,Schiminovich07}, a
region bounded in UV-optical color-magnitude---star-formation rate
(sSFR) parameter space by the traditional ``blue cloud'' [rest-frame
(NUV$-r^{\prime}$)$\simeq2$ mag; average
$\langle\,sSFR\rangle\simeq10^{8-9}$ yr$^{-1}$] and the ``red sequence''
[(NUV$-r^{\prime}$)\gsim5~mag; average
$\langle\,sSFR\rangle\simeq10^{10-12}$ yr$^{-1}$; see e.g., Mostek et
al.~2013; Barro et al.~2013].   The fuel for this star formation may be
supplied via cold-gas accretion (Lucero \& Young 2007; Serra et
al.~2012) and/or mergers (Naab, Johanssen, \& Ostriker 2009; Dekel,
Sari, \& Ceverino 2009; Kaviraj et al.~2014),  the latter being
ubiquitous in the $\Lambda$CDM paradigm of hierarchical galaxy assembly
\citep{ElicheMoral10,Khochfar03}. The rate of major mergers is too low
at intermediate redshift, thus if RSF is driven by mergers these are
likely to be minor mergers (i.e., with mass ratios, $\mu$\gsim1:4; see
Lopez-Sanjuan 2010, 2012).   

High spatial resolution imaging has also confirmed high-redshift
($z\sim$2) ETGs are systematically more compact (i.e., higher stellar
density and smaller effective radii) than are ETGs in the local universe
(Daddi et al.~2005; Cassata et al.~2011, 2013; Trujillo et al.~2006, van Dokkum et al.~2008, Damjanov et al.~2009). In
simulations, optimal size-mass growth in ETGs by mass accretion is
accomplished via dry (i.e., gas-free; see Naab, Johanssen, \& Ostriker
2009) minor mergers. Yet dry minor mergers, by definition, do not
introduce the cold gas that is necessary for star-formation.  If minor
mergers underpin both the observed size evolution of compact ETGs and
recent star-formation (RSF), the apparent implication is that the evolution of
ETGs is finely-tuned---too few wet mergers in cosmological simulations
and the observed frequency of recent, minor star-formation is difficult
to explain; too few dry mergers, and the observed size growth is not
achieved. It has been recently pointed out that gas-poor minor mergers
are not precluded from growing ETGs to their observed sizes in the local
universe, but such mergers can not yield more than $\sim$4\% of the
total stellar mass in young stellar populations \citep{Sonnenfeld14}.

To alleviate the tension surrounding the role of mergers, the loss of
baryonic mass by winds has been implicated to explain the observed
size-mass evolution (Fan et al.~2008,2010; Damjanov et al.~2009), though
this mechanism remain controversial (see, e.g., Ragone-Figueroa \&
Granato 2011). Alternatively, many have considered the possibility that
a progenitor bias affects the selection of massive ETGs at high redshift
for study, thus biasing the measurement of the size-mass evolution for
quiescent, compact ETGs (Scarlata et al.~2007, Poggianti et al.~2013;
Cassata et al.~2013).  Recently, Carollo et al.~(2013) extended such a
study to include the COSMOS survey \citep{Scoville07} and concluded that
the introduction of systematically larger ``quenched'' ETGs---the
progenitors of which were once star-forming disk galaxies---dominate
the observed size-mass evolution of ETGs since $z\sim1$.  

Thus, morphologically-selected ETGs in the ``green valley'' are likely
to be either quiescent ETGs that have moved {\it away} from the red
sequence due to the formation of young stars via mergers/cold-gas
accretion {\it or} recently-quenched, formerly disk-dominated that are
moving {\it towards} and will eventually transition to red-sequence.
Without HST's high spatial resolution and the UV sensitivity of the
WFC3, it is difficult to study recent star-formation and mechanism(s) by
which massive galaxies at intermediate redshift ($z$\gsim$>$0.5) can
evolve with respect to the ``green valley.''  At this redshift range,
minor merger remnants and stellar clusters are unresolved or undetected
\citep[][respectively]{Peirani10,Salim12}\footnote{the GALEX
point-spread function FWHM $\sim$5$\farcs$0, in comparison to HST WFC3
UVIS PSF FWHM $\sim$0\farcs1.}, and the rest-frame FUV and NUV is not
observed with SDSS \& GALEX.   In \cite{R12} we presented observations
of $\sim$ 100 intermediate redshift (0.35$<z<$1.5) ETGs made with the
Hubble Space Telescope (HST) Wide Field Camera 3 (WFC3) as part of the
Early Release Science (ERS) program \citep{Windhorst11}. The ETGs'
rest-frame UV-optical colors suggested that a large fraction of these
ETGs have likely undergone a minor burst of recent (t\lsim1 Gyr)
star-formation.  Here, we extend this work by measuring the
characteristics of the young and old stellar populations, taking a more
general approach to investigate how these ETGs may have approached the
intermediate-redshift green valley.

The outline of the paper is as follows.  In \S\ref{sec:catalog}, we
briefly describe the selection criteria \citep{R12} used to define the
ETG sample. In \S\ref{sec:sedanalysis}, we present constraints on the
age and mass of the young and old stellar populations derived from the
0.1---2.0$\mu$m SED of each ETG.  We measure the \Sersic\,profile and
the number of likely companions for each ETG in \S\ref{sec:spaceanaly},
taking advantage of the superior spatial resolution, stable PSF and low
sky-background at UV-optical-near-IR wavelengths of the HST WFC3 UVIS/IR
and ACS.  In \S\ref{sec:discussion3} we investigate correlations between
these quantitative morphological parameters and the age and mass
fraction of the best-fit young stellar populations Throughout this
paper we assume a $\Lambda$CDM cosmology with $\Omega_{m}$=0.27,
$\Omega_{\Lambda}$=0.73, and H$_{0}$=70~km s$^{-1}$ Mpc$^{-1}$
\citep{Komatsu11}. We use the following HST filter designations:~F225W,
F275W, F336W, F435W, F606W, F775W, F850LP, F098M, F125W, and F160W
represent the HST WFC3 and ACS filters. FUV and NUV represent the GALEX
150 and 250 nm filters, respectively \citep{Morrissey05}. We quote all
fluxes on the AB-magnitude system throughout \citep{Oke83}.

\section{Observations and Data}\label{sec:catalog}

Near-UV and near-IR observations were acquired as part of the WFC3 ERS
program (HST Program ID \#11359, PI:~R. W. O'Connell), a 104 orbit
medium-depth survey using the HST WFC3 UVIS and IR cameras
\citep[see][for full details]{Windhorst11}. The ERS program observed
$\sim$50 square arcminutes in the Great Observatories Origins Deep
Survey-South (GOODS-S) field \citep{Dickinson03,Giavalisco04} with the
HST WFC3 UVIS in three filters (F225W, F275W, F336W) and $\sim$ 40
square arcminutes with the WFC3 IR in three filters (F098M, F125W, F160W).
We prepared mosaic images for all UVIS and IR filters drizzled to a pixel
scale of 0$\farcs$090 pixel$^{-1}$. We re-binned the existing ACS images
(F435W, F606W, F775Ws, and F850LP) to match the pixel scale of the ERS
mosaics.

We will use the ETG catalog identified in Rutkowski et al.~(2012)
throughout the following analysis. These criteria selected ETGs that
have$\colon$

\begin{itemize}
\item been imaged in all UV and IR bands to a uniform depth in
  the ERS field;\smallskip
\item a spectroscopically-confirmed redshift measured in the range
  0.35\lsim$z$\lsim1.5;\smallskip
\item a visual morphology characterized by a centrally-peaked
  light-profile, which declines sharply with radius, a high degree of
  azimuthal symmetry, and a lack of visible internal structure which
  is characteristic of ETGs.\smallskip
\end{itemize}

Applying these selection criteria to the WFC3 ERS field, we identified
102 ETGs.


In this analysis, we will use the panchromatic (10-filter) photometry
for each ETG as measured in Rutkowski et al.~(2012). This measured
photometry for the ETGs was obtained with {\it Source Extractor}
\citep[hereafter \sex;][]{BertinArnouts96} in dual-image mode, using the
F160W image mosaic as the detection image. Previously, we determined
90\% recovery limits for simulated bulge profiles with half-light radius
of 1$\farcs$0\ equal to F225W\,\lsim\,26.5, F275W\,\lsim\,26.6,
F336W\,\lsim\,26.4, and F435W\,\lsim\,26.7 mag, respectively. We
interpret ETGs with magnitudes fainter than these recovery limits as
1$\sigma$ upper limits. We refer the reader to Rutkowski et al.~(2012)
for the complete photometry tables and details regarding the selection
and classification of the catalog ETGs. 

\section{Characterizing the Stellar Populations}\label{sec:sedanalysis}
  
\subsection{Single-Component SED Analysis}\label{subsec:onecomp}


The mean rest-frame optical \gmr~colors and optical (M$_r^{\prime}$)
absolute magnitudes measured for the majority ($>75\%$) of ETGs in
Rutkowski et al.~(2012) are in general agreement with the optical colors
observed for low-redshift red-sequence galaxies, provided a small
($\sim$0.2 mag) color correction is applied to correct for passive
evolution of the stellar population to the redshift range we considered.
This suggests that the stellar mass budgets of the ETGs are likely
dominated by relatively old, low-mass stellar objects.  The traditional
paradigm of ETG formation predicts that these stellar populations formed
in a short (t\lsim1 Gyr), massive burst of star-formation at high
redshift ($z$\gsim3; see Kaviraj et al.~2013). As a result, we expect
that the optical-IR SEDs---the wavelength regime in which stellar light
from t$\gg$1 Gyr populations dominate the spectrum of old
galaxies---should be well-described by stellar population synthesis
models that assume a comparable star-formation history (SFH). 


We produced a library of stellar population models composed of
Bruzual and Charlot (2003; hereafter, BC03) population synthesis
templates to use for characterizing the old stellar
populations in these ETGs. Star-formation in these models was
defined by a single-burst, exponentially-declining SFH with the
star-formation rate, $\psi$, characterized by
$\psi(t)\!\propto\!e^{-t/\tau}$. We calculated models for N=16 values
of $\tau$, the decay of the star-formation history, defined with a
logarithmic stepsize of:

\begin{equation}
\Delta(\mbox{log}(\tau[\mbox{Gyr}]))=\frac{max(\mbox{log}(\tau))-min(\mbox{log}(\tau))}{(N-1)}=0.28,
\end{equation} 

\noindent over the range $-2.0 <
$log$(\tau$[Gyr]$) <2.0$.  The model ages were defined over the range
1$\times10^{8} < t$(yr)$ <13.7\times10^{9}$ with a logarithmic
step-size of log($\Delta$t[yrs])$\simeq 0.02$. A Salpeter stellar
initial mass function was assumed \citep[cf.][]{vandokkum10} and the
metallicity was fixed at solar.  We applied the \cite{Calzetti00}
prescription for dust extinction, assuming 0\lsim$E(B-V)$\lsim$1$ mag
characteristic of low-redshift ETG and spheroidal galaxies
\citep{Kaviraj11}.

\begin{figure}[htbc] \begin{center}
\epsfig{file=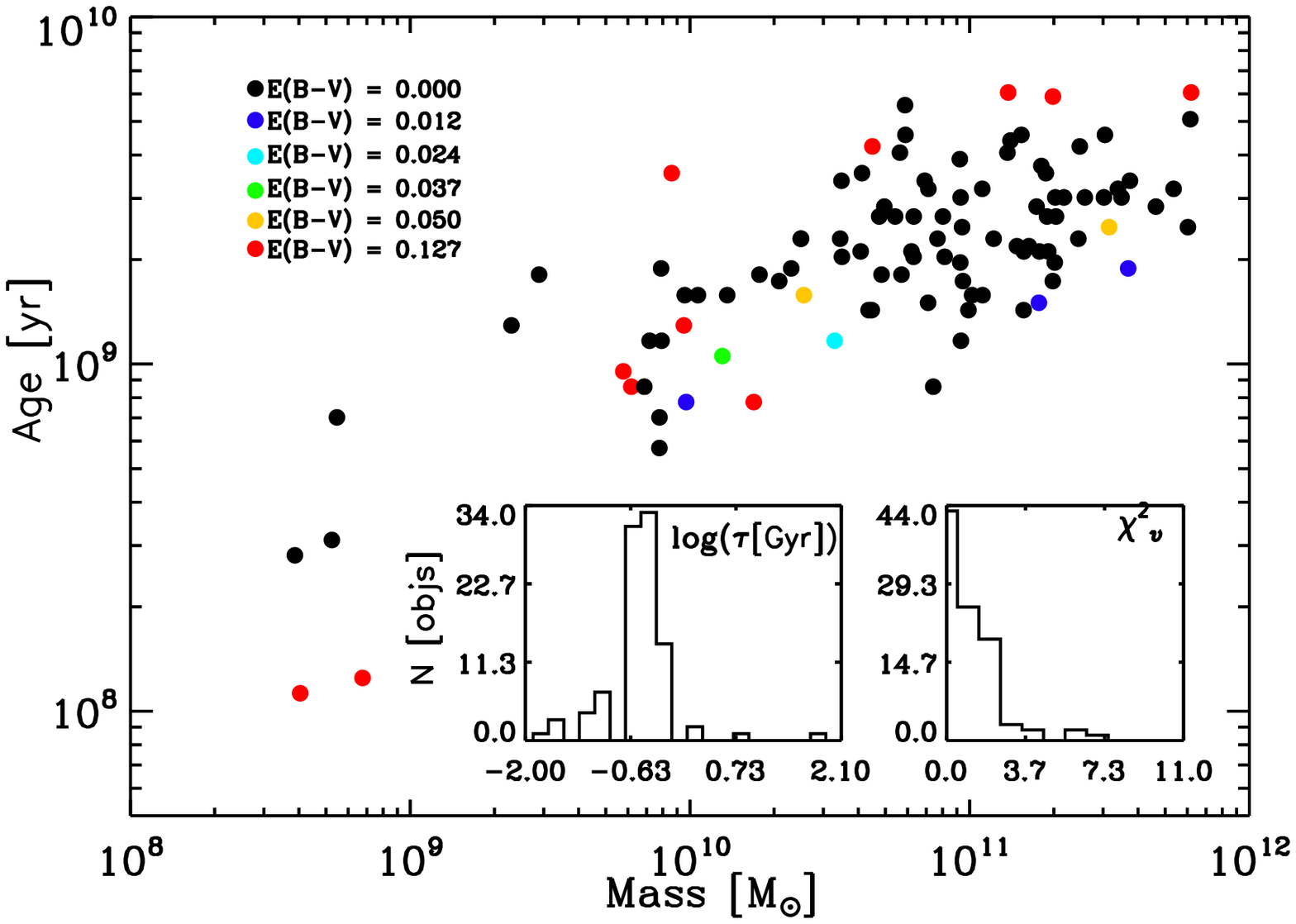, width=1.0\textwidth,clip=}
\caption{The mass (M$_{\odot}$) and age (yr) of the {\it old} stellar
populations of the ETGs, as measured from the best-fit stellar template
(\S \ref{subsec:onecomp}).  Stellar templates were fit only to the
Optical+IR SED (F435W, F606W, F775W, F850LP, F098M, F125W, F160W). In
the primary panel, we plot the measured mass-age distribution of ETGs,
coded by the best-fit dust extinction E(B-V) assuming a Calzetti (2000)
extinction law.  Inset in this panel are the distributions of the
best-fit $\tau$ (the timescale for the exponential decline of the
star-formation history, see \S\ref{subsec:onecomp}) parameter (left)
derived from the SED fits, and reduced $\chi^2$ values for each fit. In
general, the majority of ETGs selected here on visual morphology have
optical-near IR SEDs dominated by light from old stellar populations,
with little dust, and the majority of star-formation likely concluding
by $z$\gsim2.5.} \label{fig:massageonecomp} \end{center} \end{figure}

We fit these models to the broadband observed optical-IR (F435W, F606W,
F775W, F850LP, F098M, F125W, F160W) SEDs of each ETG, measuring the
best-fit model parameters by minimizing the goodness-of-fit
$\chi^2_{\nu}$ statistic for each ETGs' observed SED, following the
standard formalism of \cite{Papovich01}. 
Old, low-mass (M$\sim$\Msol) stellar populations emit predominantly at
optical-near-IR wavelengths observed with this filter set, thus
(temporarily) excluding the UV ensures that our one-component fitting is
largely insensitive to any RSF.  In the fitting process, we fix the
redshift of each ETG to its spectroscopic redshift (Table 1; Rutkowski
et al.~2012). The age of the universe at this redshift was used to set
the maximum allowable stellar age in the library of models considered in
the analysis of ETGs' SED. In Figure \ref{fig:massageonecomp}, we
present the best-fit mass and age parameter measured from these
one-component SED fits.  The SEDs of the majority of ETGs in this sample
are dominated by a massive, (M\gsim10$^{10}$\Msol), relatively dust-free, 
old (t$>2-3\times$10$^9$ Gyr) stellar population---in agreement
with the traditional paradigm of galaxy assembly for ETGs.

\subsection{Modeling Recent Star-Formation with Two-Component SED Models}\label{subsec:twocomp}

By design, the single-component SED analysis in the previous section is
sensitive to the majority (by mass) old stellar populations in these
ETGs. Here, we aim to better constrain the {\it complete} SFH, modeling
the SEDs by including synthetic stellar templates that describe both the
old and young stellar populations.  Well-studied lines and broad
absorption complexes (e.g., H$\alpha$; Ca H and K, Balmer breaks) at
optical wavelengths cannot be used here because only broadband imaging
was publicly available for the majority of these ETGs.  Enabled by the
UV sensitivity of the HST we can instead use broad-band UV diagnostics to characterize recent
(t\lsim1 Gyr) star-formation in these ETGs.  

Before characterizing any young stellar population in the ETGs, we must
first confirm that the UV emission does not likely arise from old,
low-mass (M\lsim1\Msol) stellar populations.  Such hot (T$>25000$K)
stellar populations \citep[e.g., extreme horizontal branch (EHB) stars,
see][]{OConnell99} can produce a ``UV Upturn'' (UVX). The evolution of
the EHB progenitors is not fully understood, but it is believed to be a
metallicity-dependent mass-loss effect in old (t$>$10Gyr) stars (Yi et al., 1998, 1999).

 We apply the UV-optical color-color criteria defined
 by \cite{Yi11} to differentiate between UVX-dominated and recently
 star-forming ETGs. {\it None} of the ETGs in our sample are found to
 be dominated at UV-optical wavelengths by the emission from an EHB
 population.  This agrees with the expectation from the theory of the
 evolution of low-mass stars to the EHB (see, Yi et al., 2003) that at
 $z$\gsim0.3 the UVX will be negligible as the stars in galaxies at this redshift are
 too young \citep[see,][]{Kavirajb07}. We cannot exclude the possibility
 of an EHB population, but---if they are present
 \citep[cf.][]{Han07}---this is a minority stellar population in
 low-redshift ETGs \citep[$\ll$1\% of the total stellar mass,][]{Yi11}.
 Where they exist, the blue rest-frame UV-optical colors measured in
 \cite{R12} likely indicate the presence of a relatively young (t\lsim1
 Gyr) population of massive stars.

The characteristics of any minor, young stellar population in these
optically red ETGs will not be well-constrained if the SFHs are modeled
with a library of stellar population templates that assumes
(\S\ref{subsec:onecomp}) the star-formation rate is well-described by a
{\it single}, exponentially-declining starburst event from high ($z>3-4$)
redshift.  We extend our analysis of the SFHs to include
two-component stellar template models, in which the SFH is defined by
two, independent bursts of star-formation of varying mass fractions and
ages.

In this analysis, we follow the same methodology defined in
\cite{Jeong07}.  We fit two-component synthesized stellar populations
model {\it simultaneously} to the observed ERS photometry, minimizing
the $\chi^2_{\nu}$ of the model fit to measure the best-fit model to
each ETGs' ten-filter SED.  As in Jeong et al.~(2007), we applied a
template library of stellar population models for which the old stellar
population is modeled by the Y$^2$ models which include EHB stars (Yi et
al.~2003), assuming an initial burst of star-formation at $z\simeq3$
\citep{R12,Kaviraj13}. The second component is designed to represent any
possible young stellar component and is derived from the BC03 templates
assuming a fixed solar metallicity for all models.  When fitting the
complete UV-optical-near IR SED, only the mass of the old stellar
population models was variable. In contrast, both the age (hereafter,
$t_{YC}$) and the mass fraction
($f_{YC}$=$\frac{M_{young}}{M_{total}}\times100$\%) of the younger
stellar component were variable---10$^{-3}<t_{YC}$[Gyr]$<$5 and
10$^{-4}<f_{YC}[$\%$]<$100.  Both stellar populations components could
be derived from the BC03 models in principle because the color-color
measurements demonstrate that the contribution from UV-bright, old
stellar populations  are likely to be negligible. 

If the characteristics of the young and old stellar populations are
derived from an SED analysis which uses a library of two-component
synthesized stellar population templates derived {\it exclusively} from
the BC03 templates, the derived age and mass fraction of the young
stellar populations are in good agreement ($\sim$90\% of the ETGs agree,
to within the measurement uncertainties) with those measured using
synthesized BC03$+$Y$^2$ libraries as discussed above. For the few ETGs
where they exist, we can attribute discrepancies between the best-fit
age and mass fraction measured using these two libraries to degeneracies
in the fitting arising from the large photometric uncertainties in one
or more of the bands that assesses rest-frame UV emission (i.e.,
these ETGs have a signal-to-noise ratio in the UV \lsim1)\footnote{For the
comparison made in the above text, we produced a library of models
derived exclusively from BC03 to those discussed in the text in which we
assume 1) the majority of the stellar mass in the ETGs was formed in a
short ($\tau<100$Myr) burst of star-formation at high ($z_f\sim3.5$)
redshift and 2) a secondary, more recent burst
(0.01$<t_{YC}$[Gyr]$<5.0$) with a varying mass fraction
(10$^{-3}<f_{YC}$[$\%$]$<$100). In general, the results of fitting two-component
models derived from BC03 {\it or} BC03 and Y$^2$ stellar libraries were
very good agreement, as expected}.

For ease of comparison of these results with similar measurements of the RSF in ETGs
at lower redshift, we chose to define the two-component synthesis models
used here identical to Jeong et al.~(2007).  Note, in this two-component
SED analysis, we did not apply an explicit correction for dust.  Dust
preferentially attenuates the SED at UV wavelengths, thus this fraction
of ETGs that are found to have experienced a minor, recent
star-formation event is a {\it lower} limit to the true fraction.

\begin{figure}[htbc]
\centering
\epsfig{file=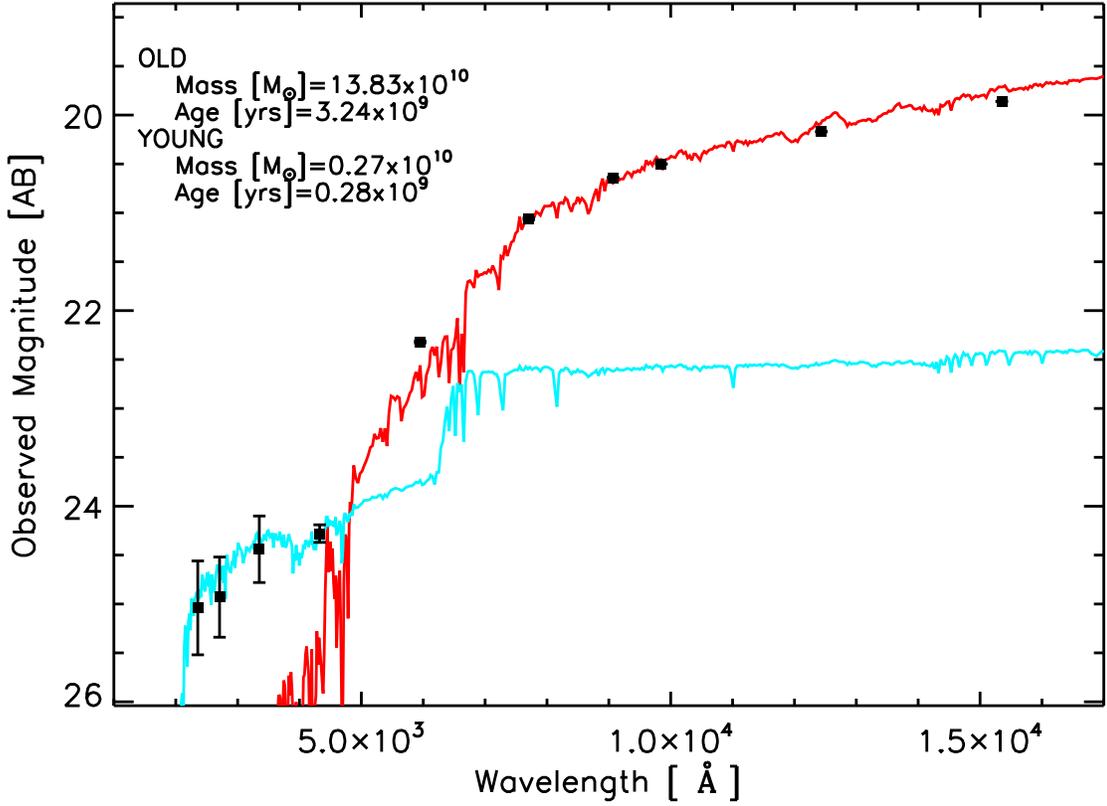, width=1.0\textwidth,clip=} 
\caption{A representative fit of the two-component synthetic stellar
  population to the ten-band SED measured for a catalog ETG
  (J033212.20-274530.1). Here, the contribution of the old stellar
  population, constituting a majority of the stellar mass in this
  galaxy, is plotted in red. The young stellar component is plotted in
  blue. Best-fit parameters associated with each stellar population
  are inset in the figure.}
\label{fig:repfits}
\end{figure}

A representative two-component model fit SED is shown in Figure
\ref{fig:repfits} for ETG J033212.20-274530.1.  In Table
\ref{tab:yspfits}, we present the best-fit parameters from our
two-component SED analysis, with upper and lower uncertainties on the
measurement of each free parameter indicating the 68\% confidence
level.  Where they exist, large young stellar population parameter
uncertainties can be primarily attributed to the \textit{photometric}
uncertainties associated with the WFC3 UVIS data. The ERS program is a
medium-depth survey and observed these UV-faint (F225W\gsim~23 mag) ETGs
to a signal-to-noise ratio in the range 1\lsim S/N\lsim20 (see Table 1
in Rutkowski et al.~2012).  These photometric uncertainties are markedly
lower\footnote{This is a testament to the improved UVIS capabilities of
the HST considering the total exposure time ($\sim$2 orbits) for the
field at UV wavelengths.} than measured in previous surveys of
comparable galaxies at this intermediate redshift range
\citep[e.g.,][]{Ferreras00}.  Although $\chi^2_{\nu}$ values of the
best-fit models are generally small ($\chi^2_{\nu}$\lsim$1-2$), we
caution that the uncertainty in the measurement $t_{YC}$ \& $f_{YC}$ are
not correspondingly small, due to the implicit degeneracies in fitting
these models to SEDs with large photometric uncertainties. Furthermore,
distinguishing between a massive old stellar population and a relatively
low-mass young (t\lsim50 Myr) starburst---in which the UV-light can be
strongly attenuated by the young stellar population dusty ``birth
cloud''---using broadband photometry alone is difficult \citep[see
e.g.,][]{Kavirajb07}.  These systematic effects are more pronounced at
the high redshift ($z>1$), a range in which the observed ERS filters are
insensitive to rest-frame wavelengths greater than $\sim$1$\mu$m, where
old stellar populations dominate the SED.

Throughout the text, we will define ETGs to have evidence for recent
star-formation when the characteristics of the best-fit young stellar
population from this two-component SED analysis is in the range of
1$<f_{YC}$[\%]$<10$ and 0.1$<t_{YC}$[Gyr]$<1$. Applying this criteria,
we conservatively measure \textit{at least} $\sim40$\% of ETGs to have
experienced a recent, minor burst of star-formation in a sample of 77
total ETGs well-fit (i.e., $\chi_{\nu}^2$) from this two-component
analysis.    The mean age and mass fraction of the best-fit young
stellar population component for the sample equals to $t_{YC}=360\pm160$
Myr and $f_{YC}=3.7\pm2$\%. At low redshift ($z$\lsim0.1),
\cite{Kavirajb07} observed $\sim$30\% of ETGs to have UV colors
consistent with recent star-formation, with an average age of the young
stellar component of $\sim$300--500 Myr.  At higher redshift (1$<z<$3),
\cite{Kavirajb13} found $\sim60$\% of massive {\it spheroidal} galaxies
have evidence of recent star-formation and to have a redshift of formation
$z_f\,\simeq\,3-4$.  Considering only the {\it magnitude} of the
observed fraction of ETGs with RSF suggests that star-formation in ETGs
generally declines with decreasing redshift.

 Each of these previous measurements of the fraction of ETGs with RSF is
 a {\it lower} limit to the total fraction at a given redshift,  as the
 measurements are subject to the photometric completeness in each
 survey, particularly at the UV wavelengths most sensitive to recent
 star-formation in these optically red ETGs. This completeness can vary
 considerably between surveys. For example, the GALEX Medium Imaging
 Survey (MIS)---which was used in the measurement of the low-redshift
 fraction of ETGs with RSF in Kaviraj et al. (2008)---is complete to
 NUV$<$23; our ERS data probes UV magnitudes 2.5 magnitudes {\it
 fainter}.  We measure a fraction of ETGs with RSF approximately equal
 to 30\% (11/38 ETGs) consistent with the measurement in the local
 universe, if {\it only} ETGs measured to have $M_{r^{\prime}}<-21.3$
 (see Table 5, Rutkowski et al. 2012) are considered (i.e., including
 only those galaxies as bright as were considered in the sample of
 Kaviraj et al.~2008).

\begin{figure}[h]
\centering
\epsfig{file=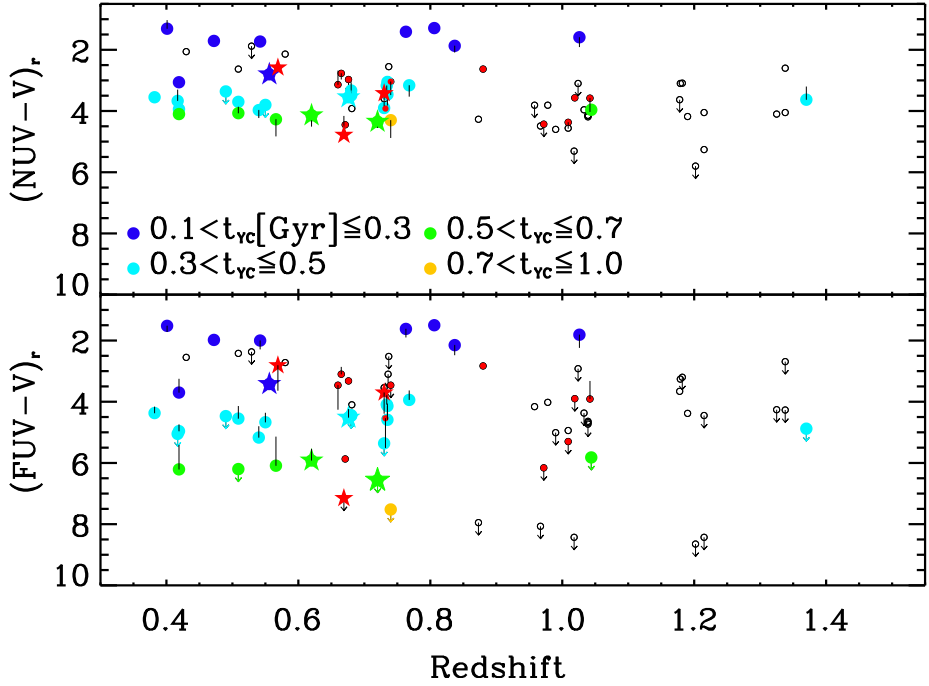, width=0.8\textwidth,clip=}
\caption{The rest-frame \nuvv (top) and \fuvv (bottom) colors of the
ETGs measured from the best-fit ($\chi_{\nu}^2<2)$ two-component model
(see \S \ref{subsec:twocomp}), plotted with a color scheme indicating
the age of the best-fit young stellar component. Large, filled points
indicate ETGs meeting conservative requirement defined in
\S\ref{subsec:twocomp} for RSF (1$<f_{YC}\mbox{[\%]}<10$;
0.1$<t_{YC}\mbox{[Gyr]}<1$).   Small, red filled points indicate ETGs
consistent with recent star-formation, but not meeting conservative
criteria for recent star-formation (see \S\ref{subsec:twocomp} for
details).  Small circles indicate ETGs best-fit with a two-component
model consistent with a quiescent star-formation history.  Throughout,
X-ray/radio sources are indicated by star symbols.  Overplotted for
these data are the offsets between the UV-optical colors derived from
the analysis in \S\ref{subsec:twocomp} and the measured colors, inferred
from the HST photometry and presented in Rutkowski et al.~(2012). These
offsets are small ($\Delta\ll$0.3; for clarity, only offsets larger than
0.2 mag are plotted) indicating the method applied in Rutkowski et al.~
(2012) to transform observed colors with the HST filter set to the
rest-frame GALEX FUV \& NUV optical colors are valid for intermediate
redshift ETGs. Data with a small, downward pointing arrow indicate that
the UV-optical color of the galaxy was reported as an upper limit in
Table 5 of Rutkowski et al.~(2012).} \label{fig:NewColors} \end{figure}

In Figure \ref{fig:NewColors} we provide the (FUV-V)$_{r}$ and
(NUV-V)$_{r}$ colors of the ETGs measured from the two-component
analysis, plotted with respect to their spectroscopic redshifts. Here,
ETGs measured with recent star-formation (as defined by the conservative
criteria above) are indicated as large, filled colored points, with an
inset color scheme corresponding to the best-fit young stellar
component.  In addition, we include those ETGs (smaller, red, filled
points) with measurement uncertainties of the age and stellar mass
fraction of young stars consistent (i.e., $<$1 dex, see
Table \ref{tab:yspfits}) with recent star-formation. Black, unfilled circles
in the figure indicate ETGs whose best-fit young stellar parameters
are consistent with a quiescent star-formation history.  In addition to
these measured data, we incorporate additional data for a subset of
these ETGs from the catalogs presented in Rutkowski et al.~(2012). We
overplot X-ray/Radio sources as filled
star symbols, and recently star-forming ETGs with UV-optical colors
previously reported as upper limits are overplotted with a small,
down(red)ward-pointing arrow. We also overplot (vertical lines) the offset
between the UV-optical color reported in Rutkowski et al.~(2012) and
those measured from the best-fit two-component model.  For
clarity, we only show the color offset when the difference between the
rest-frame colors derived from the broadband transformation in Rutkowski
et al.~(2012) and the model colors is greater than 0.2 mag. Few
($\sim15\%$) ETGs show large offsets---this confirms that the
generalized transformation of the observed broadband UV-optical colors
to the rest-frame GALEX UV-optical colors is reasonable for the majority
of intermediate redshift ETGs. 

In Figure \ref{fig:NewColors}, in particular for the (FUV-V)$_r$ colors,
there appears to be an evolution from high to low redshift towards
relatively {\it bluer} colors---in contradiction with the initial
conclusion that the sample is consistent with a general decline in the,
or potentially constant, fraction fraction of ETGs experiencing recent
star-formation from $2<z<3$ to z$\sim$0. The trend does not likely
represent a cosmological effect. Rather, the eye is strongly biased by
the few extremely red (FUV-V\gsim\,7) ETGs at $z$\gsim1 whose colors
were derived from the two-component SED analysis. In this analysis,
these ETGs are indicated with downward-pointing arrows indicating that
these ETGs were non-detected in the rest-frame FUV.  Additionally, these
galaxies are relatively faint ($m_{F606W}\sim26$) at optical
wavelengths. Uncertainties in the measurement of the best-fit model from
which these UV-optical colors are derived are further compounded by the
fact that at high redshift ($z>1$) the reddest ERS filter (F160W) is
sensitive to rest-frame optical ($\lambda<8000$\AA) emission. As a
result, at $z$\gsim1 the stellar mass in old stars becomes increasingly
difficult to constrain without the near-IR rest-frame data available for
the ETGs at lower redshifts. Similarly, the young stellar mass fraction
{\it relative to} the total old stellar mass becomes increasingly
difficult to constrain. Thus, for these ETGs---which were measured to
have a mass fraction consistent with zero---the derived UV-optical
colors will be very red, though not for physical reasons but as a result
of these compounding photometric and modeling uncertainties.  The
initial conclusion that the fraction of star-forming ETGs declines or remains
constant, as a function of decreasing redshift, remains
tenable. We note that the addition of deeper UV photometry and IR
rest-frame photometry (though existing archival data is at significantly
lower spatial resolution) could improve the uncertainties associated
with the higher redshift ETGs' rest-frame UV-optical measured photometry
and stellar population fitting.

\section{Morphological Analysis of ETGs and their Local
Environments}\label{sec:spaceanaly}

Feedback from starbursts or AGN \citep{SilkRees98} in the progenitors of
massive ETGs may have expelled or destroyed the fuel necessary for
subsequent star-bursts \citep[see simulation results
from][]{Kavirajb07,Schawinski09a,Kaviraj13}.  If quiescent ETGs are to
form new stars then they must acquire cold gas by accretion and/or minor
mergers from the local environment.  Alternatively, the ``quenching'' of
star-formation in later-type S0/lenticular galaxies
\citep{Kannappan09,Lucero13} or disk galaxies (e.g., Carollo et al.~2013)
could force these galaxies to transition away from the blue-cloud  and
towards the red-sequence as {\it in situ} gas reservoirs are consumed on
short (t$\sim$1Gyr) timescales \citep{Schawinski14}.

The high spatial resolution and continuous UV-optical-near-IR coverage
of our HST WFC3 data allows us to directly search for evidence of the
mechanism(s) driving the observed recent star-formation. In
\S\ref{subsec:sersec}, we measure \Sersic\,profiles \citep{Sersic63} of
the ETGs to determine the relative fraction with bulge- and disk-like
dominated light-profiles. A joint consideration of the quantitative
morphologies (\S\ref{subsec:sersec}) in conjunction with the
star-formation histories (\S\ref{subsec:twocomp}) and companion analysis
(\S\ref{subsec:comps}) may provide clues to the mechanism by which
ETGs formed young stars. For example, if a positive correlation exists
between the frequency of disk-like light-profiles in association with
ETGs {\it in isolated environments} (i.e., no satellites are observed in
these deep UV-optical-near IR data) in close proximity
to the ETG {\and} with evidence of RSF, this could implicate the quenching
of star-formation in (formerly) disk-dominated galaxies as an important
mechanism for motivating RSF. If this result exists in conjunction with
the measurement of a {\it negative}, for example, correlation between
the frequency of companions and RSF in ETGs then this could imply that
such ``quenching"---in contrast to environmental factors (e.g.,
mergers)---are relatively {\it more} important for motivating RSF in
intermediate redshift ETGs.

\subsection{Quantitative Morphology of ETGs}\label{subsec:sersec}
These ETGs were identified in Rutkowski et al.~(2012) by visual
selection based on their similarity with the ``classical'' morphology
of ETGs: a high degree of rotational symmetry and smoothly varying
stellar light-profile.  Such light-profiles can be well-described with
relatively few parameters, for example, the \Sersic\,profile defined
as: 

\begin{equation}
I(r)=I(0)\times\,\exp[-b_n(r/r_e)^{1/n}],
\end{equation}

\noindent where $I(0)$ is the intensity at radius $r=0$, $r_e$ is the
half-light radius, $n$ is the \Sersic\,index, and $b_n$ is a
normalization constant that is a function of the \Sersic\,index and
ensures that the radius $r_e$ encloses half of the total galaxy light.
Generally, disk-dominated galaxies are described by a \Sersic\,profile
with $n\simeq$1; bulge-dominated, spheroidal galaxies are best-fit by
\Sersic\,profile with $n\simeq4$.  A large dispersion for spheroidals is observed, though.  At low redshift, \cite{Kormendy09} measured a mean
\Sersic\,index of $n\simeq$3.8 (N=37 ETGs) with a large spread ($>35$\%
of the ETGs were measured $n>4$; $60$\% of those were measured with
$n>7$). \cite{Krajnovic13} measured ``disk-like'' \Sersic\,profiles
($n$\lsim2) for the majority of the SAURON sample of local ellipticals
\citep{deZeeuw02}. High redshift compact, quiescent ETGs also are found
to have a large dispersion in measured \Sersic\,indices---van der Wel et
al.~(2011) and Ryan et al.~(2012) report $n$\,\lsim\,2 for 30-60\% of
compact ($r_e$\lsim 1 kpc) massive (log(M[\Msol])\gsim11) quiescent
ETGs, whereas Cassata et al.~2013 and Williams et al.~2014 report large
($n>$2.5) indices for $\sim90\%$ of ETGs.  Considering the wide range of
observed morphologies of ETGs, we can reasonably expect to find a
similar ``diversity'' in this sample of visually-selected ETGs,
especially as our sample selection includes, e.g., S0s/Lenticulars  and
compact ETGs (see Table 2 in Rutkowski et al. 2012).  

A quantitative assessment of the ETGs morphologies may reveal unique
clues to the assembly histories of the ETGs. For example, in simulations
of gas-rich major mergers at high redshift, Wuyts et al.~2010 found that
the ETG descendants of such mergers have surface brightness profiles
best characterized by large \Sersic~indices ($n$\lsim~10), with a core,
young component associated with the final coalescence of the merger. In
our sample, a visual inspection of the rest-frame UV morphologies (see
Figure 1 in Rutkowski et al. 2012) reveals that for the $\sim$30\% of
the ETGs which show appreciable UV emission, this light appears to be
dominated by core emission.  The combination of the blue UV-optical
colors in these ETGs and core-dominated UV light-profiles could, in
light of this Wuyts et al. result (and others, cf. Hopkins et al. 2008,
2009) result, provide clues to formation and evolution of the ETGs. 

We used \texttt{GALFIT} \citep{Peng02} to first measure the best-fit
\Sersic\,profile for each ETG in 200kpc$\times$200kpc postage stamp
images extracted from the ERS F160W mosaics.  We implemented
\texttt{GALFIT} via the IDL software wrapper \textit{iGALFIT}
\citep{Ryan11}, which is useful for iterative batch processing.  This
software requires the user to provide an image and weight map in the
same units (counts s$^{-1}$), ensuring that the uncertainty ($\chi^2_{\nu}$)
of each profile fit in \texttt{GALFIT} is properly normalized.

We prepared large (250-500 square pixels) postage stamps for each ETG
for analysis with {\tt GALFIT}. We masked a large (20\lsim N\lsim50)
number of regions in each postage stamp that contained neighboring
galaxies or noisy pixels (e.g., at WFC3 chip and mosaic gaps).  It was
never necessary to mask more than $\sim$10\% of the total image area,
but this masking is necessary. \texttt{GALFIT} calculates the sky
brightness locally within each image, and fits the model light-profile
assuming that all flux within the region of interest is associated with
the ETG. Identifying and removing the contaminating sources (e.g.,
foreground and background objects) ensures a more accurate measurement
of the light-profile.

In Table 2 of Rutkowski et al.~(2012), 15 galaxies were noted for their
compact morphologies, so it is also necessary to exclude ETGs that may
be only marginally spatially-resolved. We identify marginally-resolved
ETGs to exclude by fitting all ETGs with a \Sersic\,profile and an
empirical PSF \citep[defined by stacking known stars in the ERS field,
see][]{Windhorst11}.  We then calculated the fractional $\chi^2$
difference, $F_{crit}$, equal to:

\begin{equation}
  F_{crit}=\frac{(\chi^2_{PSF}-\chi^2_{\mbox{\Sersic}})}{\chi^2_{\mbox{\Sersic}}}, 
\end{equation}

\noindent where $\chi^2$ is measured from the two model fits
\citep{Bond09}.  \cite{Ryan12} determined that for ERS ETGs observed in
F160W, $F_{crit}\simeq$0.01 can generally distinguish point-sources from
well-resolved galaxies.  13 galaxies were measured to have
$F_{crit}<0.01$ and are designated ``\textit{Failed $F_{crit}$}'' in
Table \ref{tab:yspfits}.  Nine of these ETGs were originally noted for
their ``compact'' morphology in Rutkowski et al.~2012 and a visual
inspection of publicly-available spectra\footnote{available online at
\texttt{http://archive.eso.org/archive/adp/GOODS/FORS2\_spectroscopy\_v3
.0/index.html}} confirmed that $\sim$50\% (7/13) those ETGs were
identified with [OII]3727\AA, or an unknown emission line, in their
spectrum potentially indicating the presence of central star cluster or
a weak AGN.  If the stellar light-profiles of these ETGs was relatively
faint in comparison to a bright spatially-unresolved point source, this
could explain the poor \Sersic\,profile fit. We exclude these galaxies
from the subsequent analysis. At this stage, we also exclude one
additional ETG because the light-profile of this galaxy was inextricably
blended with a nearby galaxy and no accurate mask model could be
determined.  This ETG is indicated ``\textit{Not Fit}'' in Table
\ref{tab:yspfits}.

Next, we inspected the residual images produced by \texttt{GALFIT} by
differencing the best-fit \Sersic\,model light-profile and the original
input image. We found $\sim$20\% ETGs that were not excluded by
the above criteria were still poorly fit by a single
\Sersic\,profile\footnote{If \texttt{GALFIT}~ can not converge on a
  parameter solution after a finite number of iterations, it will
  designate the poorly constrained parameter with an asterisk, ``*''.
  The reduced $\chi^2$ for the model fit may be small (\lsim~1), but
  this solution should not be considered robust.}.  These images
typically showed irregular, patchy or ``ring''-like structure (a
bright core, bounded by an over-subtracted region) in their residual
map. This structure may indicate the presence of an
additional component, either in the core \citep[a ``cuspy'' core or
  centrally-concentrated star-forming region, see e.g.,][]{Suh10} or
wings (possibly indicating an extended core stellar-disk component).

To improve our \Sersic\,model fit, we re-measured the light-profiles
using a two-component model composed of a combined \Sersic\,model and
the empirically-defined PSF.  We repeated our fitting of the
light-profiles with this model, measuring the best-fit model ($M$)
which satisfied the criteria that 1) \texttt{GALFIT}~parameter solution converged for each model and 2) 
$M=\min\{\chi^2_{\mbox{\Sersic}},\chi^2_{\mbox{\Sersic}+PSF}\}$, where
the $\chi^2$ was reported by \texttt{GALFIT}. In Table
\ref{tab:yspfits}, ETGs whose light-profiles
were better modeled with this two-component spatial model are
designated with $\chi^2_{\nu}$ in a boldface font.  No solution could be found for 2
ETGs with either the one (\Sersic\,only) or two (\Sersic\,$+$empirical
PSF) component model. Rather than attempt a fit with additional
components, we designate the row value for these ETGs as
``\textit{Fail to Converge}'' in Table \ref{tab:yspfits} and do not
consider these ETGs in the following discussion. In summary, 86 of 102
ETGs were well-fit ($\langle\chi^2_{\nu}\rangle$=0.54) with either the
one- or two-component \Sersic\,model. 

\begin{figure}[h]
\centering
\epsfig{file=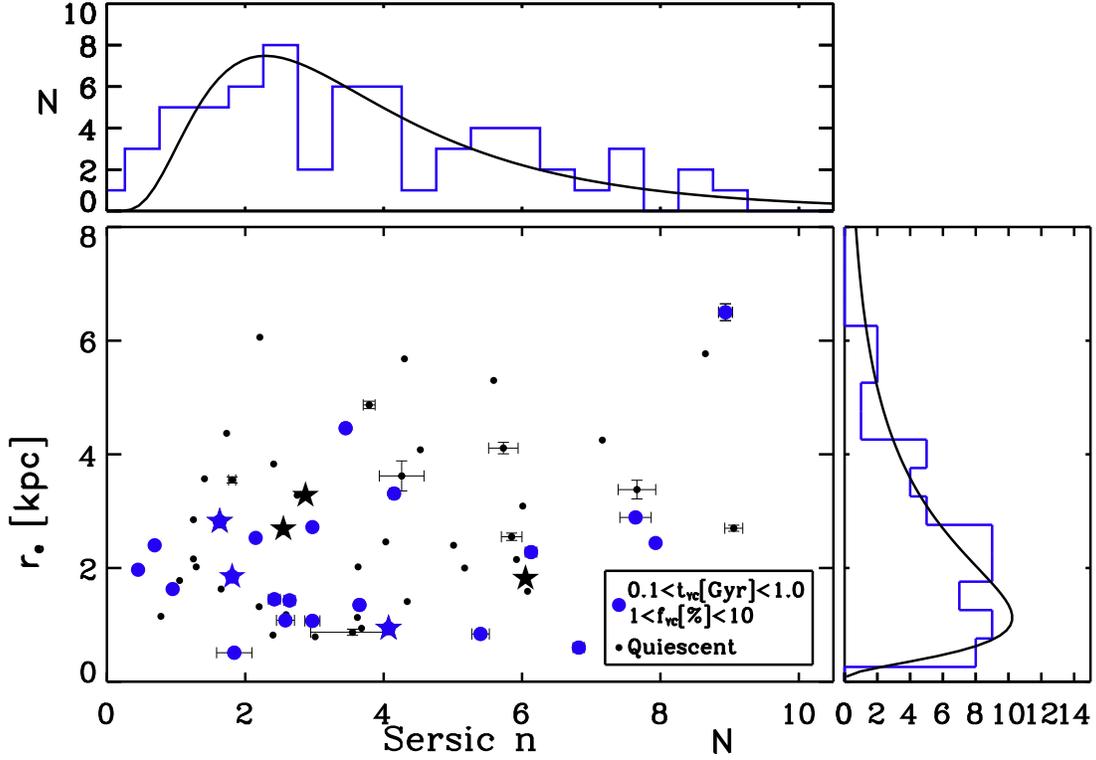, width=0.9\textwidth,clip=} 
\caption{In \S \ref{subsec:sersec}, we measured the best-fit
  \Sersic\,function with index $n$, effective radius $r_e$ and
  ellipticity for the F160W light-profile of each ETG; here, \Sersic\,
  $n$ is plotted against $r_e$ for ETGs well-fit ($\chi<2$) in this
  analysis. Color of data points distinguish those ETGs identified
  with recent star-formation from quiescent ETGs, well-fit
  ($\chi^2<2$; N=73 ETGs---20 with RSF \& 53 quiescent,
  respectively) in the morphological analysis. For clarity, we only
  overplot measurement uncertainties larger than $1\%$. ETGs
  identified with an AGN are designated with a filled star, with a
  color indicating the best-fit model. We fit a log-normal function to
  the distribution \Sersic\,index of half-light radii and plot these
  in the top and right hand panels.  The best-fit mean \Sersic\,index
  and half-light radii, $\langle\,n\rangle$=2.1 \&
  $\langle\,r_e\rangle$=1.4.}
\label{fig:NvsREage}
\end{figure}

In Figure \ref{fig:NvsREage}, we plot the best-fit half-light radii
(converted to a physical scale of kiloparsecs, assuming the
spectroscopic distance) against the measured \Sersic\,index, with the
symbol colors indicating the age of the young stellar population,
$t_{YC}$.  The mean \Sersic\,index $\langle n \rangle$ equals
$3.7\pm2.1$ and the mean half-light radius of $\langle {r}_e\rangle$
equals $2.9\pm1.88$\,kpc. In the top panel of Figure
\ref{fig:NvsREage}, we plot a log-normal function fit to the
distribution of $n$, with $\langle\!n\rangle$=2.1, $\sigma$=1.2 and
skewness, $\gamma$, equals 1.6. In the right panel of Figure
\ref{fig:NvsREage}, again we plot a log-normal function fit to the
distribution of $r_e$, with a best-fit mean, variance (in kpc) and
skewness equal to 1.4, 1.3, and 2.0, respectively.   In general, there
is no strong correlation observed between characteristics of the young
stellar populations identified from the analysis of
\S\ref{sec:sedanalysis} and the morphology of these ETGs, i.e., recent
star-formation is observed for ETGs with both disk ($n<2.5$) and
bulge-like ($n>2.5$) morphologies. We note that the mean \Sersic~index
for high-mass ETGs (M$>10^{10.5}$\Msol) equals $\langle\!n\rangle$=4.2
($\sigma$=2.2); the mean index measured for low-mass ETGs equals $\langle\!n\rangle$=2.2
($\sigma$=1.2).  For a discussion of the possible implications of these
results for the evolution of ETGs at intermediate redshift, we refer the
reader to \S\ref{sec:discussion3}.

\subsection{ETGs and Likely Companions}\label{subsec:comps}

Mergers of companions with massive ETGs are implicated in theory and
observation to explain the recent star-formation observed in the
latter.  The space density of {\it major}
($\mu=\frac{M_{comp}}{M_{ETG}}\sim1$, i.e., equal mass) mergers is not
high enough to account for the observed star-formation in intermediate
to high-redshift ETGs (e.g., Lopez-Sanjuan 2010,2012; Kaviraj et al.~2013).
Minor ($\mu\geq$1:4) mergers occur more frequently than major mergers,
and are as a result a more viable mechanism for introducing gas into ETGs
\citep{Kavirajb13}.  Due to the relative short ``destruction''
timescales of companions in minor mergers \citep{Peirani10}, catching
such mergers ``in the act'' in order to directly correlate recent
mergers with the incidence of recent star-formation is not feasible
with this small sample. Instead, the presence of companion galaxies
can be used as a proxy for future mergers as the infall time are
markedly longer ($\gg$1 Gyr; Tal et al.~2013).  In the following
sections, we outline (\S\ref{subsec:statmeth}) and apply
(\S\ref{subsec:compnomeas}) a method for measuring the number of {\it
  likely} companions for each ETG.

\subsubsection{Likely Companions to ETGs: Method}\label{subsec:statmeth}

In simplest terms, we want to identify galaxies in close proximity to
each ETG.   To do this, we define a ``volume of interest'', {\it V},
centered on each ETG with dimensions (on the plane of the sky) of
$\{X,Y\}= \{X_{ETG}\pm100$ kpc$, Y_{ETG}\pm100$ kpc$\}$, centered on the
ETG. The third spatial dimension is proportional to the relative
velocity of a galaxy in this volume. Here we adopt
$v_{comp}=v_{spec,ETG}\pm 750$~km s$^{-1}$. The dimensions of this
volume are comparable to the definition applied in studies of pairs of
galaxies at similar redshifts \citep[e.g.,][]{Ryan08,LopezSanjuan10}. 
We specifically select this volume because the likelihood of a merger of
the ETG and the companion(s) by z$\simeq$0 is predicted to be greater
than 50\ from simulations \citep{Tal13}.

Not all galaxies located in this search region are likely companions,
though. If the three-dimensional spatial positions of the galaxies
within this region are well-constrained by a spectroscopic redshift
then measuring the companion number to each ETG is a simple counting
exercise and the number of likely companions equals exactly to the
number of galaxies in the search region. In practice, identifying
likely companions is more difficult, since both high-resolution
imaging \textit{and} complete spectroscopic data are not available for
all possible companions to ETGs in the ERS field, despite extensive
efforts.  In Rutkowski et al.~(2012) we estimated the spectroscopic
redshift deficit for visually-classifiable (i.e., early or late-type)
galaxies in the ERS field, and the fraction
\textit{without} measured spectroscopic redshifts may be as large as
$\sim$75\%.  This spectroscopic incompleteness arises from technical
limitations. First, spectroscopic redshift campaigns are limited by
the apparent brightness of the observed galaxies.  The
\cite{Vanzella08} spectroscopic survey of this field, for example, is 
likely to be only $\sim$10-20\% complete for the faintest galaxies
(F850LP$>$25 mag) including ETGs. Note that this spectroscopic
incompleteness also implies a mass incompleteness for the catalog of 
potential companions. Secondly, quiescent ETGs, which lack significant
line-emission, may be undetected as the Ca 4000\AA~broad absorption
complex can not be bracketed from the ground at the high redshift
range for this sample.  Third, the ground-based spectroscopic
candidate selection is typically done in the $i$-band ($i$\lsim\,22-24
mag), but for the reddest $z$\gsim\,1 ETGs, our H-band selection
selects additional candidates for which spectroscopic redshift can not
be well-constrained due to the atmosphere.

We measured photometry in all ten HST ACS and WFC3 filters for all
galaxies in stamps, centered on the ETG, with an area equal to
$\pi\times(100 \mbox{kpc})^2$ (equivalently, stamp sizes with widths of
250-500 pixels at 0.35\lsim$z$\lsim1.5) using \sex~ in dual-image mode.
The F160W image was used as the detection image, and we applied the
\sex~detection criteria outlined in Rutkowski et al.~(2012). After
excluding objects that were located on or near the edges of the search
region, we fit the observed photometry in all filters, using
the {\tt EAZY} \citep{Brammer08} software to measure the galaxies'
 photometric redshifts. The SEDs were fit with all combinations of the
 SED templates\footnote{These spectral templates are derived from the
 P\`{E}GASE model SEDs \citep{Fioc97} by the authors of the {\tt EAZY}
 software} provided with {\tt EAZY} . We refer the readers to the {\tt
 EAZY} manual\footnote{available online at
 http://www.astro.yale.edu/eazy/eazy\_manual.pdf} for full details
 regarding this spectral template library, but note that these models
 generally represent the range of SFHs for galaxies at intermediate
 redshifts.  We then produced a list of {\it possible} companions within
 a broad redshift range ($z_{ETG}\pm$0.25) for each ETG.  We matched the
 photometric redshift catalogs of all galaxies with existing
 spectroscopic redshift catalogs for the GOODS-S field from the
 literature, where available, to produce a catalog of {\it possible}
 companions for each ETG. Possible companions that were identified in
 both catalogs were assigned the higher-precision spectroscopic
 redshift. Few companions in each stamp were measured with
 spectroscopic redshifts (N$<$3, at most).

We further reduced the catalog of possible companions by applying a
magnitude selection, requiring the F160W magnitude measured for the
possible companion to meet the criterion
$m_{F160W,comp}<m_{F160W,{ETG})}+2.5$. If the stellar mass-to-light
ratios of these galaxies are similar, this implies that the stellar
mass ratio of the pairs are 1\lsim$\mu$\lsim10.

We apply a probabilistic method for measuring $N_c$, the number of
likely companions, motivated by the formalism defined first in
Lopez-Sanjuan et al.~(2010).  This method formally incorporates the
measurement uncertainty in the photometric redshift of a galaxy into the
measurement of $N_c$.  This is
accomplished by assuming that the likelihood of identifying a galaxy in
the volume of interest as a companion to the ETG is proportional to the
quality of the redshift. Specifically, if a galaxy has a spectroscopic
redshift, the redshift probability function (or more generally, the
``PF") defining the likely position of a galaxy with respect to the
volume of interest is the Dirac delta function, equal to one within the
volume and zero elsewhere. If a galaxy has a photometric redshift,
Lopez-Sanjuan et al.~assume that its PF is Gaussian. The {\tt EAZY}
software reports a unique redshift PF for each galaxy, which we adopt
instead for possible companions that do not have spectroscopic
redshifts.   To measure the number of likely companions, we 1) integrate
the probability of each system of galaxies---in each integration, one
member is always fixed to be the ETG---to be a pair and 2) weight by the
sum of the probabilities that each galaxy exists in the volume of
interest to measure the weighted pair probability.  We then sum over all
weighted pair probabilities for all possible pairs to define $N_c$. For
complete details on this method, we refer the reader to Appendix A.

\subsubsection{Likely Companions to ETGs: Results}\label{subsec:compnomeas}

Applying the Lopez-Sanjuan formalism we find that the total number of
likely companions (N$_c$, defined as the sum of the individual system
contributions, $\nu_k$, as outlined in the Appendix) is only greater than one {\it if} the galaxies
within the volume of interest are measured with a {\it
spectroscopic} redshift. Conversely, {\it no} ETGs were measured with
$N_c$\gsim1 if {\it all} possible companions had {\it only} measured
photometric redshifts. For systems---with possible companions measured
only with photometric redshifts---to contribute $\nu_k\simeq1$, the
uncertainty in the companions' redshift must be quite small.  Only in
the idealized case where the photometric redshift of the possible
companion equals to the spectroscopic redshift of the ETG, will a
Gaussian PF with $\sigma_z$\lsim$10^{-3}$ (or, equivalently an
uncertainty in velocity equal to $\sigma_{v}$\lsim$10^2$~km s$^{-1}$)
contribute $N_{c}^{j}\sim$1. No possible companions with photometric
redshifts and such narrow PFs met the selection criteria
(\S\ref{subsec:statmeth}).  In practice, the PFs associated with
photometric redshifts for possible companions were measured with
$\sigma_z\simeq10^{-1}$.  Uncertainties of this magnitude imply contributions
to the total number of likely companions in these pairs of
\lsim10$^{-2}$.  The total number of galaxies with only photometric
redshifts (i.e., no companions considered had {\it spectroscopic}
redshifts) was never greater than 7 in the search region.  Thus, the
cumulative contribution of these system was never greater than
$N_{c}\simeq0.1$. Increasing the search volume could increase the number
of possible companions considered. Doing so will also decrease the
likelihood for a merger by z$\sim$0 below $\sim$50\% based on the
results of Tal et al.~2013, though. We can not expect that the PFs of
the possible companions will be appreciably improved by the use of more
broad or medium band filters. In the zCOSMOS survey, for example,
\cite{Knobel12} used imaging in 30 medium- and broad-band filter to
measure photometric redshifts and found even this extensive coverage to
be insufficient for identifying individual pairs and groups for galaxies
at $z$\lsim\,1.

We have measured more than one likely companion for $\sim$20\% (16/102)
of the ETGs only where the companions' redshifts were spectroscopically
confirmed.  As such, this measurement is a strong {\it lower} limit on
the frequency of companions to ETGs. In Table \ref{tab:tablex}, we
present a list of these ETGs.  We include 1$\sigma$ uncertainties
associated with $N_c$, measured with an empirical jackknife technique
following \cite{Efron82}, where$:$ \begin{equation}
\sigma^2=(N-1)\sum\limits_{j=1}^N \frac{(N_c^i - N_c)^2}{N},
\end{equation}

\noindent here $N_c^i$ is the total number of galaxies excluding the
contribution from the $i^{th}$ system, $N_c$ the total number of likely
companions, $N$ equals the total number of galaxies, and the sum is
measured over the set of {\it j} possible companions. In Table
\ref{tab:tablex}, we also include the number of possible companions with
photometric (Col. 3) and spectroscopic (Col. 4) redshifts which were
contributed to measurement of $N_c$.

We measure the average characteristics for galaxies with and without
companions for the subset of ETGs (N=33) measured with recent
star-formation (1$<f_{YC}\mbox{[\%]}<10$; 0.1$<t_{YC}\mbox{[Gyr]}<1$) in
\S\ref{subsec:twocomp} and with morphological parameters well-fit
($\chi^2_{\nu}<$2).  For ETGs with $N_c<1$ (N=26), the mean \Sersic\,index,
mass, and sizes equal
$\langle\,n,\,$M[\Msol]$,\,r_e$[kpc]\,$\rangle\simeq(3.6\pm$2.5,\,10.5$
\pm$0.59,\,2.6$\pm$1.7), with $1\sigma$ dispersion reported on each mean
value. The mean age and mass fraction of young stars for this subset
equals
$\langle\,t_{YC}$[Myr],\,$f_{YC}$[\%]$\rangle\simeq(325\pm$185,\,3$\pm$2
). For the sub-sample of ETGs with N$_C\geq1$ (N=6),
$\langle\,n,\,$M[\Msol]$,\,r_e$[kpc]$\rangle\simeq(3.9\pm$1.5,\,10.7$\pm
$0.52, 2.3$\pm$1.9). The mean age and mass fraction of young stars for
this subset equals $\langle\,t_{YC}$[Myr]$,f_{YC}$[\%]$\rangle\simeq$
(260$\pm$130,\,2$\pm$1.9). If these data are normally distributed in
each sample, a two-sample t-test shows that the mean parameters
presented for ETGs with, and without, companions are not statistically
significant.  Whether these measurements are in fact normally
distributed is difficult to determine given the small sample sizes, but
we conclude that the measured properties of ETGs with and without
companions are indistinguishable. 


\section{The size-mass relation for intermediate redshift ETGs}\label{sec:discussion3}

It is useful to frame the results from our analysis in the context of
the physical size and stellar mass (i.e, ``size-mass'') relationship of
galaxies, a touchstone for both theoretical and observational studies of
ETG evolution. In the following sections, we present this measured
bivariate distribution for our intermediate redshift ETGs
(\S\ref{subsec:discussion3a}).  We discuss the measured distributions in
the context of previously-published size-mass relationships and the 
implications for the evolution of intermediate redshift ETGs in \S\ref{subsec:discussion3b}. 

\subsection{The observed size-mass relation}\label{subsec:discussion3a}
\begin{figure}[htbc]
\begin{center}
\epsfig{file=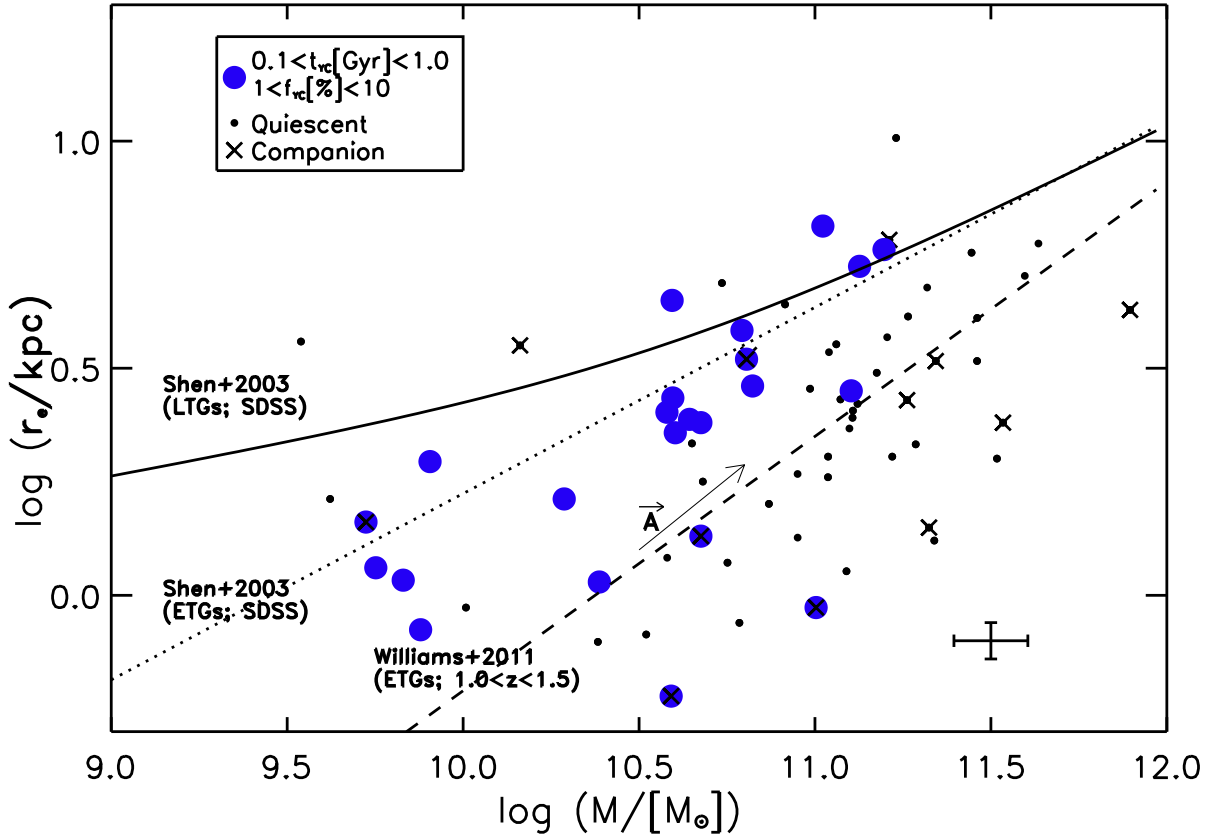,width=1.0\textwidth,clip=} 
\caption{We summarize the results of the SED modeling, morphological and
companion analyses here, plotting the size and stellar mass for all ETGs
well-fit (i.e., $\chi^{2}_{\nu}<2$) in both the SED modeling
(\S\ref{sec:sedanalysis}) and morphological (\S\ref{sec:spaceanaly})
analyses (N=73).  The representative mean uncertainty in the size and
mass are provided in the bottom right. Quiescent ETGs identified from
the SED fitting are plotted as small, black points; ETGs with likely
companions (\S \ref{sec:spaceanaly}) are indicated by a black
``$\times$''. Note that only those galaxies with
spectroscopically-confirmed redshifts are measured to be likely
companions to the ETGs, using the statistical method outlined in
\S\ref{subsec:statmeth}. ETGs identified in \S \ref{subsec:twocomp}
analysis with evidence of recent star-formation ($f_{YC}<$10\%; t$<$1
Gyr) are indicated with blue, filled circles. For reference, we overplot
an empirical intermediate-redshift size-mass relationship for ETGs
(dashed; Williams et al.~2010). We overplot the Shen et al.~(2003)
low-redshift size-mass relationships measured for early-type (dotted) and
late-type (solid) galaxies.  Note, massive (M$>10^{10.5}$\Msol) ETGs
with recent star-formation appear to be loosely clustered near the Shen
et al.~low-redshift ETG size-mass relationship, where as quiescent ETGs,
particularly those with companions, appear to have smaller average sizes
and cluster towards the high-redshift relationship of Williams et al.
(2010).  We overplot a preferred vector ($\vec{A}$), indicated by the
thin-line arrow originating at log(M,r$_e$)$\simeq$(10.5,0.1), which
appears to bisect those ETGs with RSF and those with $N_c>1$.  We use
the perpendicular distance of these data measured with respect to this
vector in \S\ref{subsec:discussion3b} to test whether this apparent
clustering is statistically significant.} \label{fig:SizeMassRel}
\end{center} \end{figure}

In Figure \ref{fig:SizeMassRel}, we plot the half-light radii of the
ETGs (\S\ref{subsec:sersec}) against the stellar masses derived from the
optical-IR SED fits (\S\ref{subsec:onecomp}). Here, we only plot ETGs
that were well-fit ($\chi^2_{\nu}<$2) in the analyses of {\it all}
sections.  For reference, we plot the mean uncertainties in half-light
radius and stellar mass for an ETG with an average size and mass.  We
overplot three empirical size-mass relationships observed for local ETGs
and late-type galaxies (dotted and solid black curves, respectively;
reproduced from Shen et al.~2003) and intermediate redshift
(1.0$<z<$1.5) ETGs (dashed line; from Williams et al.~2010)---these
relationships are {\it not} fits to the data.  We indicate ETGs with
likely ($N_c$\gsim1) companions within $\Delta v=750$~km s$^{-1}$ with
an ``$\times$'' symbol. ETGs that were best-fit (\S\ref{subsec:twocomp})
with a minor component of young ($1$\%$<f_{YC}$[\%]$<10$\% and
$0.1.<t$[Gyr]$<1$) stars are plotted as filled points with a color
defined by the key in the figure. ETGs not meeting these additional
criteria are plotted as small, filled circles.

A few trends appear in Figure \ref{fig:SizeMassRel} that may provide
some insight to the mechanism(s) inducing RSF in intermediate redshift
ETGs. First, there does not appear to be a mass dependency in the
distribution of ETGs that have experienced recent star-formation (t$<1
$Gyr). Though many of the low-mass galaxies (M$<10^{10.5}$) are likely
to have experienced a minor burst of RSF in the previous $\sim$1 Gyr,
RSF is observed in high-mass ETGs as well.   Secondly, we note that
high-mass (M$>10^{10.5}$\Msol) ETGs with RSF and those (albeit few) ETGs
with likely companions ($N_c\ge1$) {\it appear} to occupy unique sectors
in the size-mass parameter space. Recently star-forming ETGs {\it
appear} to be distributed on or near the Shen et al.~(2003) low-redshift
size-mass relation.  In contrast, quiescent ETGs---in particular, ETGs
with companions---appear on or near the intermediate-redshift size-mass
relationship for ETGs.   The tight apparent clusters, at M$\simeq$10.5
and 11\Msol, within the broader mass-size distributions are populated by
ETGs over a wide redshift range (0.6$<z<$1.4), though we re-iterate the
important note from Section \ref{subsec:twocomp} that these ERS data are
most sensitive to recent star-formation at the lower ($z$\lsim1)
redshift range of our samples. 

We confirm with a two-sample t-test that the means of each of these
distribution are distinguishable (i.e., the null hypothesis is rejected
at \gsim95\%, if the data are normally distributed). To better
illustrate this apparent clustering, we collapse the bivariate size-mass
distributions of these two populations of ETGs along a preferred vector
that is approximately parallel to the empirical size-mass relationships
for high-mass ETGs (Williams et al. 2010). This vector---a {\it
non-unique} bisection that was estimated from a visual inspection of the
distributions of these two populations---is overplotted (indicated
$\overrightarrow{A}$ in Figure \ref{fig:SizeMassRel}) for reference in
Figure \ref{fig:SizeMassRel} as a thin-line arrow originating at
log(M,r$_e)\simeq$10.5,0.1.  For each of the ETGs with RSF and with
likely companions we measure the magnitude of the perpendicular
distance, $\overrightarrow{dA}$ from the preferred vector
$\overrightarrow{A}$. A histogram of these distances is plotted in black
($N_c\ge1$) and red (RSF), respectively, in Figure 6. Additionally in
Figure 6, the best-fit Gaussian function to each distribution is
overplotted in the same color. Here, the mean for all quiescent
high-mass ETGs are distinguishable (null hypothesis is rejected at
$>99.5$\% level) from ETGs with recent star-formation, with the median
distinguished at $5\sigma$. If the ETGs with RSF (ETGs with$N_c\ge1$)
that lie above (below) the vector $\overrightarrow{A}$ are removed for
the purposes of refining the fit of the Gaussian to each distribution,
the means of the two distributions can be distinguished at $7\sigma$. 

\subsection{Implications for the evolution of ETGs from their size-mass
distribution}\label{subsec:discussion3b}

The observed diversity in the colors and SFHs of our ETGs is difficult
to reconcile with models which would predict more uniformly passive
characteristics for such galaxies. For example, Peng et al.~(2010)
presented a model in which massive (M\gsim$10^{10}$\Msol) galaxies
reside in a ``mass-quenching'' regime, in which these galaxies'
evolution is dominated by internal feedback, in contrast to
environmental factors. This model predicts 65--80\% of massive ETGs in
our sample will reside on the red sequence. We do not observe such a
mass-dependency for star-formation in our sample.  

The lack of strong mass-dependent trend of observed RSF may be
interpreted as support for star-formation in ETGs as a {\it stochastic}
process (e.g., RSF is related to the environment). Though the number of
ETGs with companions is small (N=10), the distinction in the
distributions may still be instructive of the process(es) driving recent
star-formation or transforming the sizes of massive ETGs. We note that
the companion galaxies to the $\sim20$\% of ETGs identified with
companions (\S\ref{subsec:compnomeas}) are bright (m$_{F606W}$\lsim23
mag) and therefore massive\footnote{This is a systematic effect, as
brighter galaxies are more likely to have spectroscopically-confirmed
redshifts and thus have been identified by analysis in
\S\ref{subsec:compnomeas}, which limits the range of mass ratios we
consider to $\mu$\gsim1:2}, and have blue UV-optical rest-frame colors (
$\langle(NUV-V)_{r}\rangle\simeq2.5$ mag).  This suggests these
companions to be relatively gas-rich.  From simulations, these
companions will likely merge (\gsim50\%) by z$\sim$0 (Tal et al.~2013).
If these companions do not consume their cold-gas reserves in advance of
a future merger then some degree of new star-formation should be
expected in the ETGs. 

Alternatively, the measured size-mass distribution and the observed
characteristics of ETGs with RSF could be interpreted as evidence for a
``progenitor bias.''  In effect, in our sample we may caught
recently-quenched galaxies ``in the act'' as they transition towards the
red sequence.  For a comparison with high ($z\sim1.5$) redshift studies,
consider the recent results of Bedregal et al. (2010, hereafter ``B10'')
which selected $\sim$40 passive galaxies from the HST WFC3 IR grism
pure-parallel WISP survey (PI:M. Malkan). Differences in the selection
of galaxies between the B10 sample and our own makes a direct comparison
difficult --- B10 used rest-frame optical color and mass to select
relatively more massive ($\langle\,M\,\rangle\simeq10^{11}$\Msol)
passive galaxies for study, whereas our selection is primarily
morphological and does not by design {\it preclude} minor recent
star-formation --- but the B10 results are germane to this discussion.
First, BC10 also observe a diversity in the average ages of the stellar
populations in massive, quenched galaxies, with approximately 30\% of
the galaxies residing off of the passive, red sequence. They interpret
the ``homogeneous spread'' in the mass of the quenched galaxies on and
off of the red-sequence as evidence that multiple mechanisms are
responsible for the quenching of star-formation in high mass galaxies,
and the authors emphasize that the spectra of these galaxies imply that
these galaxies are transitioning from a former period of intense star
formation at high (z$>>$2) redshift. B10 also predict that massive
galaxies in their sample would join the red sequence by z$\sim$1,
implying that young stellar populations in quenched intermediate
redshift ETGs could still be directly detectable in our broadband UV
data.

Furthermore, at low-redshift ($z$\lsim0.1), Wyder et al.~(2007) have
suggested a ``continuum'' of young and old transitioning galaxies
populate the UV-optical low-redshift ($z$\lsim0.1) green valley.  In our
sample, amongst massive ($>$10$^{10.5}$\Msol) recently star-forming ETGs
we do find disk-like ($n<$2.5) ETGs, which supports an interpretation
that the observed RSF could be associated with the
quenching of star-formation in higher redshift progenitors of local
ETGs.  We note, though, the mean \Sersic\,index measured for these
high-mass ETGs (see \S\ref{subsec:sersec}) is large
($\langle\,n\rangle\simeq$4.2), and is similarly large for both the
sample of recently star-forming and quiescent high-mass ETGs
($\langle\,n\rangle\simeq$4.5).  It is important to note that the high
mean \Sersic~need not imply that the histories of massive galaxies have
followed a relatively quiescent evolution.  Carollo et al.~(2013), in
the study of thousands of $z\sim$1 ETGs found that $\sim$90\% of high
mass (M$>10^{10.5}$\Msol) star-forming disk galaxies are bulge-dominated
in their light-profiles.


We observe 1) the association of companions with ETGs, albeit a small
fraction of the total number of ETGs in the catalog, 2) the presence of
bulge-dominated profiles in both compact, quiescent and recently
star-forming ETGs, 3) recent star-formation independent
of ETG stellar masses, and 4) a distinction between the size-mass
distributions of ETGs with RSF and ETGs with likely companions. 
Considered jointly, these observations imply that both the introduction
of quenched galaxies as well as mergers both likely play a
non-negligible role in the formation of young stars and size-mass
evolution of intermediate redshift ETGs. 

This conclusion is most severely limited by the relatively small number
of ETGs identified with possible companions that have {\it
spectroscopic} redshifts.  We have obtained spectroscopic observations
of an additional $\sim$100 intermediate redshift ETGs and possible
companions in the COSMOS field with MMT Hectospec in order to expand the
sample size of ETGs and their possible companions at \rsrang. This
survey program specifically targeted both previously unobserved bright
(F160W$<21$ mag) ETGs, as well as their bright companions. Repeating the
companion analysis presented in \S\ref{sec:spaceanaly} with the
increased spectroscopic redshift statistics--combined with deep U-band
observations ($<$27.5mag) we have obtained at the Large Binocular
Telescope---will improve the observational constraints of the role of
relatively gas-rich minor mergers in the mass-size evolution and stellar
mass-assembly of ETGs.

\section{Conclusion}

We used HST WFC3 panchromatic data to study the mechanism(s) that induce
the observed recent star-formation, here confirmed in {\it at least}
40\% of the intermediate redshift (\rsrang) ETGs.  This measurement is
bounded by similar measurements of the fraction of ETGs with RSF
observed at high (60\%; e.g., Kaviraj et al. 2014) and low (30\%; e.g.,
Kaviraj et al. 2007) redshift.  Together, these measurements suggest
that the frequency of RSF in ETGs generally declines with decreasing
with redshift.  We caution a strict interpretation of this result,
noting that each of these measurements is limited by the photometric
completeness (particularly at UV wavelengths) of the surveys from which
the fraction of RSF is observed. We can not rule out a constant fraction
of ETGs with RSF with respect to decreasing redshift from $z\sim1.5$ to
the local universe.

We find evidence for RSF in ETGs best-fit by disk ($n<2.5$) and bulge-like
($n>2.5$) \Sersic\,profiles, with the mean \Sersic~index increasing only 
weakly with mass. Furthermore, the prevalence of RSF
in ETGs does not correlate with the mass of the galaxy.  This result is
at odds with a pure ``mass-quenching'' model of massive galaxy evolution
(cf. Peng et al.~2010), which has been postulated for the low-redshift
($z<0.1$) evolution of galaxies.

In addition, we find that massive (M$>10^{10.5}$\Msol) ETGs with
evidence of RSF appear to be large on average and cluster towards the
low-redshift size-mass relationship for ETGs measured by Shen et
al.~(2003). Quiescent, massive ETGs have smaller sizes, but with a
larger dispersion than is measured for ETGs with RSF.  This result
suggests that the introduction of recently-quenched star-forming
galaxies into the green valley and the red-sequence, may motivate the
observed RSF and size-mass evolution observed since $z\sim2$ (Carollo et
al.~2013) .  We can not rule out---due to the relatively small number of
likely companions with confirmed spectroscopic redshifts---an
environmental, gas-rich minor merger scenario that induces RSF in
intermediate redshift ETGs. Considering the frequency of companions to
ETGs with RSF, a large dispersion in the size-mass distributions and the
presence of bulge- and disk-like light-profiles measured for quiescent
and recently star-forming ETGs, we suggest that both the transition of
disk-like progenitors to the red sequence {\it and} minor-merger/recent
interactions are both important in the evolution of intermediate
redshift ETGs. Future deep, large volume UV-optical surveys, in
combination with deeper spectroscopic surveys that precisely measure the
redshifts of the faint potential companions, will be ideally suited to
differentiate the relative roles of environmental (e.g., minor mergers)
and progenitor bias in motivating the observed frequency of RSF in
intermediate redshift ETGs.


We note that the confirmation of recent star-formation in ETGs at this
redshift range required a significant investment in space-based,
rest-frame UV-optical observations.  We urge the community in future
surveys of the star-formation histories of massive galaxies at
intermediate redshift to include deep UV rest-frame observations prior
to the decommissioning of the HST in the coming decade.

\section{Acknowledgements} 

This paper is based on Early Release Science observations made by the
WFC3 Scientific Oversight Committee.  We are grateful to the Director of
the Space Telescope Science Institute, Dr. Matt Mountain, for generously
awarding Director's Discretionary time for this program. Finally, we are
deeply indebted to the crew of STS-125 for refurbishing and repairing
HST.  Support for program \#11359 was provided by NASA through a grant
from the STScI, which is operated by the Association of Universities for
Research Inc., under NASA contract NAS 5-26555.  MR acknowledges support
from a US State Department Fulbright Junior Research Fellowship in
conjunction with the S. Korean government and Yonsei University. HJ
acknowledges support from Basic Science Research Program through the
National Research Foundation of Korea (NRF) funded  by the Ministry of
Education (NRF-2013R1A6A3A04064993).  MR thanks C. Lopez-Sanjuan for
extended helpful discussions. We thank an anonymous referee for
extensive comments that have improved this publication.

\newpage
\appendix
\centerline{Appendix A: Companion Probability}

Numerous techniques exist for counting pairs and groups of galaxies for
which photometric and/or spectroscopic redshifts have been measured. The
distinctions between these methods arises from the treatment of the
uncertainties ($\frac{\Delta\,z}{1+z}$ = 5-10\%) in the
three-dimensional positions of the galaxies.  In the simplest scenario,
when the positions of galaxies with spectroscopically-confirmed
redshifts are measured with high precision in high spatial-resolution
images determining the number of companions to any given galaxy is a
counting exercise.  In deep broadband multi-wavelength surveys, the
positions of the majority of galaxies are typically constrained only by
photometric redshifts from broadband SED fitting. When the uncertainties
on these redshifts are small, these measurements may still prove very
useful though in measuring statistical trends for large samples of
galaxies (e.g., the measurement of the pair fraction of galaxies or the
merger rate) by incorporating the positional uncertainty as a weight in
calculating the likelihood of two galaxies being companions.

Lopez-Sanjuan et al.~(2010) presented a statistical study of the pair
fraction of galaxies at intermediate redshift, defining the number of
pairs distributed amongst {\it k} systems at a redshift $z_1$ as:

\begin{equation}
\nu_k(z_1) = C_k P_1(z_1|\eta_1)\int^{z_{m}^+}_{z_{m}^-} {P_{2}(z_2|\eta_2)} dz_2
\label{eqn:ncequation}
\end{equation}

\noindent where $C_k$ is a constant normalizing the number of pair
systems to unity, $P(z|\eta)$ the probability function (or, simply the
PF referenced first in \S\ref{subsec:statmeth}), and [$z_{m}^-,z_{m}^+$]
is the redshift range of interest.  To measure $\nu_k$, we define each
of these terms as follows.  The redshift range [$z_{m}^-,z_{m}^+$] is
defined for some range over which pairs are cosmologically meaningful;
functionally, this equals to \{$z_{\rm m}^{-}$,$z_{\rm
m}^{+}$\}=$\{z_{ETG}\times$(1-$\frac{\Delta\,v}{c}$)-$\frac{\Delta\,v}{c
}$, $z_{ETG}\times$(1+$\frac{\Delta\,v}{c})$+$\frac{\Delta\,v}{c}\}$.  
We assume $\Delta\,v$=750 km s$^{-1}$, a range motivated by Tal et al.
(2014) to include close pairs that will merge with a probability greater
than $50\%$ by $z\sim$0. Incorporating the spatial positions of the
possible companions, we can define the volume of interest, {\it
V}=$\{X,Y,Z\}$ proportional to $\{X_{ETG}\pm100$ kpc$, Y_{ETG}\pm100$
kpc,$v_{comp}=v_{spec,ETG}\pm 750$~km s$^{-1}\}$.  In practice,
Lopez-Sanjuan define the probability function, $P(z|\eta)$, of
identifying a galaxy in the redshift range with respect to the quality
of its measured redshift. If a galaxy is measured with a spectroscopic
redshift $z_s$, the probability of finding it within the range $\{z_{\rm
m}^{-},z_{\rm m}^{+}\}$ is given as $P_s(z|\eta)= \delta(z - z_s)$,
where, $\delta$ is the Dirac delta function.  If a galaxy in proximity
to the volume of interest has a measured photometric redshift, $z_p$,
its probability is defined as
$P_p(z|\eta)=\frac{1}{\sqrt{2\pi}\sigma_{z_{{\rm
p}}}}\exp\left\{{-\frac{(z_{ETG}-z_{{\rm p}})^2}{2\sigma_{z_{{\rm
p}}}^2}}\right\}$.  We use the EAZY photometric redshift software to
calculate redshifts for galaxies within 100kpc of each ETG, and thus
$P(z|\eta)$ equals to the PF reported by EAZY for the best-fit model fit
for each galaxy. Finally, $C_k$ is functionally equivalent to
$N^k_p=\int^{z_{m}^+}_{z_{m}^-} P_1(z_1|\eta) dz_11
+\int^{z_{m}^+}_{z_{m}^-} P_2(z_2|\eta) dz_2$. Summing over $\nu(z)$
over all k-systems in all redshift intervals of interests yields the
total number of likely companions (i.e.,
$N_c=\sum\limits_{k}\nu_k(z_1)$). In \S\ref{subsec:compnomeas} we apply
this methodology for calculating likely companions to the ETGs within
the volume of interest, fixing the the redshift range
([$z_{m}^-,z_{m}^+$]) constant for all {\it k} pair systems potentially
associated with each ETG.

As discussed in Section 4.2.2, we found that only those systems in which
the possible companion galaxies were measured with spectroscopic
redshifts were measured to have $N_c>$1.  This does not imply that the
ETGs which had possible companions with only measured photometric
redshifts had no likely companions.  Instead, this stems directly from
the PF curves associated with these photometric redshifts being too
poorly constrained, implying that their contribution to $N_c$ never
greater than $\sim$ 0.05, or similarly the probability of finding a
galaxy in the volume of interest too small ($\ll1\%$).

For reference, the specific calculation we have made here based on the
Lopez-Sanjuan formalism may be easily conceptualized in terms of the
joint probabilities that {\it any} galaxies in the volume of interest
will be identified as an ETG companion. From the probability of sets,
the cumulative union of probabilities of {\it N} independent events can
be written in the standard notation as$:$

\begin{equation}
\label{eqn:fullsys}
\begin{split}
P(\bigcup_{i=1}^N E_x)=\sum_{\substack{i_1,i_2,\cdots,i_x
    \\ 1\leq\,i_1\leq\,i_2\leq\cdots\leq\,i_x\leq\,N}}P(E_{i_1}\cap\,E_{i_2}\cdots\cap\,E_{i_x})
\end{split}
\end{equation}

As an example, when {\it N}=3, then $P(\bigcup_{i=1}^3 E_i)$ measured
for three {\it independent} events E$_A$ {\it or} E$_B$ {\it or}
E$_C$ is:

\begin{equation}
\label{eq:p1}
\begin{split}
P(E_A\cup\,E_B\cup\,E_C)&=P(E_A)+P(E_B)+P(E_C)\\
           &- (P(E_A\cap\,E_B)+P(E_A\cap\,E_C)+P(E_B\cap\,E_C))\\
           &+ P(E_A\cap\,E_B\cap\,E_C)\\
\end{split}
\end{equation}

Here, an event $E_i$ is defined as the likelihood of a galaxy in the
volume of interest to be identified as a companion to an ETG. The limits on the
union of probabilities are 0$\leq$P$(\bigcup_{i=1}^N {\rm E}_x)\leq$1.
In the case of sets of galaxies that include small numbers of possible
companions, each with poorly-constrained (i.e., broad) PFs,
P$(\bigcup$E$_i)\rightarrow$0. Alternatively, in sets with small numbers
of possible companions which have well-constrained PFs (the case for
galaxies measured with spectroscopic redshifts) or sets that include
{\it numerous} possible companions with poorly-constrained PFs
(potentially the case for groups of galaxies with well-constrained
photometric redshifts) then P$(\bigcup$E$_i)\rightarrow$1.  Using a
similar notation to Lopez-Sanjuan et al., the probability of a galaxy to
be located in the volume of interest, {\it V}, can then be defined as
P(E$_k$)=P($z_{ETG}$)$\times\int_V$(PF), with the functional form of PF
either the Dirac delta function (for galaxies with spectroscopic
redshifts) or a Gaussian or other function as discussed previously for
galaxies with photometric redshifts.

Of course, this method will not improve the probability that any
individual ETG galaxy system will be considered a pair, as the PFs are
fixed by the quality of the data.  But, if a uniform brightness is
applied in defining systems of possible companion systems (as we did in
\S\ref{subsec:comps}), Equation \ref{eqn:fullsys} provides a more
generalized extension of the Lopez-Sanjuan formalism.    When we apply
the more general method to our data, the measurement of the number of
likely companions is identical to the measurement of $N_c$ measured
using the Lopez-Sanjuan formalism presented in
\S\ref{subsec:compnomeas}.  We again find that {\it no} ETGs are
identified with likely companions using this alternative method, if the
galaxies within the volume of interest are measured only with
photometric redshifts.  Only possible companion galaxies with
spectroscopically-confirmed redshifts are found to be likely companions
to these catalog ETGs.

\clearpage
\newpage
\begingroup;
\tiny 
\begin{longtable}{lcccccccccc}
\caption{Measured Characteristics of Catalog ETGs}\\
\hline \hline 

\multicolumn{1}{c}{} &
\multicolumn{3}{c}{Stellar Characteristics} &
\multicolumn{1}{c}{} &
\multicolumn{6}{c}{Spatial Characteristics} \\ \cline{2-4} \cline{6-11}
\multicolumn{1}{l}{GOODS ID} &
\multicolumn{1}{c}{$t_{YC}$[Gyr]} &
\multicolumn{1}{c}{$f_{YC}/100$\%} &
\multicolumn{1}{c}{$\chi^2_{\nu}$} &
\multicolumn{1}{c}{} &
\multicolumn{1}{c}{$n$} &
\multicolumn{1}{c}{R$_{e}$} &
\multicolumn{1}{c}{$B/A$} &
\multicolumn{1}{c}{$\theta$} & 
\multicolumn{1}{c}{m$_{F160W}$} &  
\multicolumn{1}{c}{{$\chi^2_{\nu}$}} \\ 
 \hline \smallskip
\endfirsthead

\multicolumn{11}{c}{Table 1: Young Stellar Populations, cont.}\\ 
\hline \hline 
\multicolumn{1}{c}{} &
\multicolumn{3}{c}{Stellar Characteristics} &
\multicolumn{1}{c}{} &
\multicolumn{6}{c}{Spatial Characteristics} \\ \cline{2-4} \cline{6-11}
\multicolumn{1}{l}{GOODS ID} &
\multicolumn{1}{c}{$t_{YC}$[Gyr]} &
\multicolumn{1}{c}{$f_{YC}/100$\%} &
\multicolumn{1}{c}{$\chi^2_{\nu}$} &
\multicolumn{1}{c}{} &
\multicolumn{1}{c}{$n$} &
\multicolumn{1}{c}{R$_{e}$} &
\multicolumn{1}{c}{$B/A$} &
\multicolumn{1}{c}{$\theta$} & 
\multicolumn{1}{c}{m$_{F160W}$} &  
\multicolumn{1}{c}{{$\chi^2_{\nu}$}} \\ 
 \hline \smallskip
\endhead

\\ \hline
\multicolumn{11}{l}{Continued on next page ...}\\
\endfoot
\hline \hline 

\multicolumn{11}{l}{\textbf{Notes-} We present spatial and stellar
  characteristics measured for the ETGs in Sections
  \ref{subsec:twocomp} and \ref{sec:spaceanaly}, respectively. Columns
  2-4}\\ 
\multicolumn{11}{l}{provide characteristics of the best-fit young stellar
  population. Uncertainties associated with these parameters represent
  the}\\
 \multicolumn{11}{l}{68\% confidence interval. Cols. 5-10 provide
  quantitative characteristics of the ETGs morphology. Row values associated}\\
\multicolumn{11}{l}{ with the spatial parameters in these columns are defined as follows: ``{\it Failed
    F$_{crit}$}''=Galaxies that failed the criterion for}\\
\multicolumn{11}{l}{identifying
  well-resolved galaxies were not fit (see
  \S\ref{subsec:sersec}); ``{\it Not
    Fit}''=Galaxies were not fit because the light-profiles of the ETG}\\
\multicolumn{11}{l}{ were strongly blended with bright
  neighbors;``{\it Fail to
    Converge}''=One or more parameters could not be well-fit by {\textit GALFIT}.}\\ 
\multicolumn{11}{l}{ The $\chi^2_{\nu}$ values of ETGs in Column 10 are
  provided in {\bf bold} for the ETGs that were better fit with a two-component}\\
\multicolumn{11}{l}{  spatial model (i.e., PSF and \Sersic\,model, see \S
  \ref{subsec:sersec}.} 
 \endlastfoot

J033202.71-274310.8&0.641$_{0.000}^{0.078}$&0.580$_{0.240}^{0.420}$&7.84&&5.78$\pm$0.007&10.24$\pm$0.023&0.79$\pm$0.000&-69.3$\pm$0.07&17.18$\pm$0.00&1.020\\\medskip
J033203.29-274511.4&0.143$_{0.053}^{0.059}$&0.050$_{0.024}^{0.030}$&1.08&&\multicolumn{6}{c}{\textit{Failed F$_{crit}$}}\\\medskip
J033205.09-274514.0&0.114$_{0.023}^{0.047}$&0.058$_{0.022}^{0.082}$&0.96&&2.42$\pm$0.086&1.45$\pm$0.018&0.92$\pm$0.011&27.8$\pm$5.62&22.48$\pm$0.01&0.441\\\medskip
J033205.13-274351.0&0.114$_{0.023}^{0.029}$&0.094$_{0.038}^{0.186}$&1.21&&2.58$\pm$0.130&1.08$\pm$0.021&0.86$\pm$0.007&34.9$\pm$2.83&23.61$\pm$0.03&{\bf 0.355}\\\medskip
J033206.27-274536.7&0.571$_{0.443}^{1.929}$&0.009$_{0.008}^{0.290}$&1.10&&2.55$\pm$0.023&2.69$\pm$0.008&0.29$\pm$0.000&-89.1$\pm$0.05&21.95$\pm$0.01&{\bf 0.443}\\\medskip
J033206.48-274403.6&0.047$_{0.047}^{0.208}$&0.000$_{0.000}^{0.006}$&0.83&&1.25$\pm$0.027&2.85$\pm$0.017&0.81$\pm$0.004&27.6$\pm$0.88&21.94$\pm$0.01&{\bf 0.468}\\\medskip
J033206.81-274524.3&0.003$_{0.002}^{0.014}$&0.000$_{0.000}^{0.000}$&2.35&&2.88$\pm$0.148&1.99$\pm$0.033&0.34$\pm$0.004&-34.0$\pm$0.27&22.98$\pm$0.03&{\bf 0.401}\\\medskip
J033207.55-274356.6&0.404$_{0.349}^{0.402}$&0.030$_{0.029}^{0.970}$&0.88&&4.15$\pm$0.064&3.31$\pm$0.051&0.85$\pm$0.005&-72.6$\pm$1.35&21.31$\pm$0.01&0.400\\\medskip
J033207.95-274212.1&0.181$_{0.109}^{0.141}$&0.009$_{0.006}^{0.020}$&0.56&&3.01$\pm$0.098&0.79$\pm$0.008&0.55$\pm$0.007&68.5$\pm$0.67&22.29$\pm$0.00&0.526\\\medskip
J033208.41-274231.3&0.404$_{0.177}^{0.236}$&0.016$_{0.010}^{0.022}$&0.34&&6.82$\pm$0.083&0.60$\pm$0.002&0.91$\pm$0.003&-17.0$\pm$1.33&20.42$\pm$0.00&0.721\\\medskip
J033208.45-274145.9&0.003$_{0.002}^{0.042}$&0.000$_{0.000}^{0.000}$&1.04&&5.92$\pm$0.059&2.15$\pm$0.020&0.86$\pm$0.003&74.3$\pm$0.75&20.21$\pm$0.00&0.596\\\medskip
J033208.53-274217.7&0.072$_{0.050}^{0.089}$&0.001$_{0.001}^{0.003}$&1.95&&2.87$\pm$0.029&3.28$\pm$0.012&0.66$\pm$0.001&-87.3$\pm$0.16&22.03$\pm$0.01&{\bf 0.817}\\\medskip
J033208.55-274231.1&0.003$_{0.002}^{0.078}$&0.000$_{0.000}^{0.002}$&1.08&&4.26$\pm$0.323&3.62$\pm$0.264&0.90$\pm$0.016&43.8$\pm$7.56&24.60$\pm$0.05&{\bf 0.671}\\\medskip
J033208.65-274501.8&1.015$_{0.110}^{0.419}$&0.260$_{0.120}^{0.740}$&1.19&&1.29$\pm$0.016&2.02$\pm$0.006&0.63$\pm$0.001&-55.7$\pm$0.18&21.35$\pm$0.01&{\bf 0.603}\\\medskip
J033208.90-274344.3&0.286$_{0.031}^{0.074}$&0.340$_{0.180}^{0.660}$&1.09&&1.65$\pm$0.058&1.63$\pm$0.012&0.62$\pm$0.006&-23.8$\pm$0.74&24.70$\pm$0.01&{\bf 0.377}\\\medskip
J033209.09-274510.8&0.114$_{0.033}^{0.047}$&0.072$_{0.024}^{0.068}$&1.31&&1.84$\pm$0.255&0.51$\pm$0.013&0.40$\pm$0.027&17.0$\pm$1.964&23.79$\pm$0.01&0.368\\\medskip
J033209.19-274225.6&0.571$_{0.368}^{1.329}$&0.022$_{0.018}^{0.338}$&1.02&&1.63$\pm$0.018&2.82$\pm$0.009&0.62$\pm$0.001&-34.4$\pm$0.18&22.46$\pm$0.01&{\bf 0.551}\\\medskip
J033210.04-274333.1&0.001$_{0.000}^{0.089}$&0.000$_{0.000}^{0.000}$&0.86&&7.16$\pm$0.031&4.25$\pm$0.033&0.85$\pm$0.002&-76.6$\pm$0.38&19.71$\pm$0.00&\multirow{3}{*}{0.556}\\\medskip
J033210.12-274333.3&0.227$_{0.226}^{1.573}$&0.002$_{0.002}^{0.277}$&1.42&&4.34$\pm$0.059&1.41$\pm$0.011&0.54$\pm$0.003&51.2$\pm$0.27&21.26$\pm$0.01&\\\medskip
J033210.16-274334.3&0.001$_{0.000}^{1.699}$&0.000$_{0.000}^{0.089}$&0.82&&5.17$\pm$0.046&2.00$\pm$0.017&0.93$\pm$0.003&87.6$\pm$1.44&20.68$\pm$0.01&\\\medskip
J033210.76-274234.6&0.360$_{0.105}^{0.044}$&0.012$_{0.007}^{0.006}$&0.59&&1.76$\pm$0.002&4.11$\pm$0.005&0.73$\pm$0.000&77.8$\pm$0.10&20.85$\pm$0.00&{\bf 3.786}\\\medskip
J033210.86-274441.2&0.143$_{0.112}^{0.143}$&0.007$_{0.005}^{0.012}$&0.24&&3.68$\pm$0.201&0.94$\pm$0.024&0.47$\pm$0.007&-73.0$\pm$0.69&23.86$\pm$0.06&{\bf 0.442}\\\medskip
J033211.21-274533.4&0.052$_{0.048}^{0.108}$&0.000$_{0.000}^{0.002}$&3.67&&2.25$\pm$0.060&1.34$\pm$0.021&0.46$\pm$0.002&-30.6$\pm$0.15&24.93$\pm$0.46&{\bf 0.439}\\\medskip
J033211.61-274554.1&0.023$_{0.022}^{0.158}$&0.000$_{0.000}^{0.002}$&1.77&&5.01$\pm$0.071&2.40$\pm$0.021&0.37$\pm$0.002&-55.1$\pm$0.15&24.11$\pm$0.05&{\bf 0.407}\\\medskip
J033212.20-274530.1&0.360$_{0.074}^{0.093}$&0.024$_{0.010}^{0.018}$&1.06&&3.38$\pm$0.047&2.81$\pm$0.034&0.63$\pm$0.004&-26.6$\pm$0.43&19.89$\pm$0.01&8.33\\\medskip
J033212.31-274527.4&0.360$_{0.133}^{0.149}$&0.026$_{0.014}^{0.026}$&1.28&&\multicolumn{6}{c}{\textit{Fail  To Converge}}\\\medskip
J033212.47-274224.2&0.453$_{0.167}^{0.451}$&0.032$_{0.016}^{0.088}$&0.75&&0.45$\pm$0.019&1.97$\pm$0.020&0.85$\pm$0.007&-14.4$\pm$2.16&21.78$\pm$0.00&{\bf 0.637}\\\medskip
J033214.26-274254.2&0.181$_{0.090}^{0.074}$&0.028$_{0.018}^{0.038}$&2.37&&\multicolumn{6}{c}{\textit{Failed  F$_{crit}$}}\\\medskip
J033214.45-274456.6&0.005$_{0.004}^{0.250}$&0.000$_{0.000}^{0.013}$&1.30&&1.81$\pm$0.052&3.55$\pm$0.055&0.66$\pm$0.005&-50.9$\pm$0.85&22.09$\pm$0.01&{\bf 0.426}\\\medskip
J033214.65-274136.6&0.001$_{0.000}^{0.019}$&0.000$_{0.000}^{0.000}$&0.97&&5.73$\pm$0.213&4.11$\pm$0.101&0.88$\pm$0.006&-32.2$\pm$2.24&23.00$\pm$0.02&{\bf 0.449}\\\medskip
J033214.68-274337.1&0.161$_{0.047}^{0.066}$&0.054$_{0.030}^{0.086}$&2.84&&2.64$\pm$0.080&1.43$\pm$0.013&0.40$\pm$0.007&-29.2$\pm$0.43&22.31$\pm$0.01&0.401\\\medskip
J033214.73-274153.3&0.404$_{0.118}^{0.236}$&0.042$_{0.018}^{0.058}$&0.31&&5.40$\pm$0.127&0.84$\pm$0.010&0.56$\pm$0.006&83.2$\pm$0.54&21.44$\pm$0.00&0.647\\\medskip
J033214.78-274433.1&0.025$_{0.024}^{0.296}$&0.000$_{0.000}^{0.011}$&1.21&&1.05$\pm$0.030&1.78$\pm$0.012&0.70$\pm$0.002&-82.2$\pm$0.53&23.168$\pm$0.02&{\bf 0.425}\\\medskip
J033214.83-274157.1&0.404$_{0.118}^{0.167}$&0.046$_{0.022}^{0.054}$&0.21&&0.69$\pm$0.017&2.40$\pm$0.015&0.88$\pm$0.005&-27.2$\pm$1.70&22.055$\pm$0.00&{\bf 0.531}\\\medskip
J033215.98-274422.9&0.404$_{0.083}^{0.049}$&0.078$_{0.036}^{0.062}$&1.06&&3.45$\pm$0.057&4.46$\pm$0.030&0.55$\pm$0.002&64.0$\pm$0.19&23.704$\pm$0.03&{\bf 0.413}\\\medskip
J033216.19-274423.1&0.286$_{0.106}^{0.118}$&0.024$_{0.012}^{0.022}$&0.39&&4.43$\pm$0.757&4.11$\pm$0.535&0.81$\pm$0.033&48.6$\pm$6.00&22.920$\pm$0.05&{\bf 4.375}\\\medskip
J033217.11-274220.9&0.020$_{0.016}^{0.006}$&1.000$_{0.996}^{0.000}$&3.78&&\multicolumn{6}{c}{\textit{Failed  F$_{crit}$}}\\\medskip
J033217.12-274407.7&0.453$_{0.251}^{0.562}$&0.024$_{0.017}^{0.116}$&0.25&&2.97$\pm$0.109&1.07$\pm$0.013&0.31$\pm$0.005&-2.6$\pm$0.37&23.43$\pm$0.03&{\bf 0.386}\\\medskip
J033217.14-274303.3&0.286$_{0.031}^{0.035}$&0.040$_{0.014}^{0.018}$&0.59&&4.07$\pm$0.019&0.94$\pm$0.001&0.87$\pm$0.001&-49.8$\pm$0.40&19.72$\pm$0.00&0.533\\\medskip
J033217.49-274436.7&0.360$_{0.133}^{0.093}$&0.044$_{0.026}^{0.032}$&1.09&&8.94$\pm$0.100&6.50$\pm$0.148&0.99$\pm$0.003&-31.1$\pm$113.40&20.14$\pm$0.01&0.387\\\medskip
J033217.91-274122.7&0.013$_{0.012}^{0.077}$&0.000$_{0.000}^{0.000}$&1.51&&2.56$\pm$0.041&2.64$\pm$0.016&0.93$\pm$0.002&-59.3$\pm$1.72&22.82$\pm$0.01&{\bf 0.439}\\\medskip
J033218.31-274233.5&0.255$_{0.074}^{0.066}$&0.009$_{0.004}^{0.008}$&3.84&&5.25$\pm$0.014&3.43$\pm$0.010&0.48$\pm$0.000&-10.9$\pm$0.04&19.09$\pm$0.00&0.665\\\medskip
J033218.64-274144.4&0.001$_{0.000}^{0.003}$&0.000$_{0.000}^{0.000}$&1.51&&3.55$\pm$0.605&0.87$\pm$0.054&0.76$\pm$0.023&-15.1$\pm$5.11&23.53$\pm$0.04&{\bf 0.419}\\\medskip
J033218.74-274415.8&0.321$_{0.094}^{0.132}$&0.014$_{0.007}^{0.014}$&0.67&&2.97$\pm$0.029&2.72$\pm$0.010&0.58$\pm$0.001&61.4$\pm$0.12&22.59$\pm$0.01&{\bf 0.450}\\\medskip
J033219.02-274242.7&0.072$_{0.070}^{0.214}$&0.001$_{0.001}^{0.010}$&1.50&&2.21$\pm$0.044&6.06$\pm$0.099&0.59$\pm$0.005&17.5$\pm$0.55&22.87$\pm$0.01&{\bf 0.930}\\\medskip
J033219.48-274216.8&0.321$_{0.118}^{0.132}$&0.016$_{0.009}^{0.012}$&0.65&&7.93$\pm$0.038&2.44$\pm$0.015&0.71$\pm$0.001&-82.7$\pm$0.14&19.29$\pm$0.00&0.560\\\medskip
J033219.59-274303.8&0.404$_{0.083}^{0.105}$&0.036$_{0.016}^{0.028}$&1.23&&7.64$\pm$0.226&2.89$\pm$0.039&0.92$\pm$0.003&-36.4$\pm$1.33&22.05$\pm$0.03&{\bf 0.681}\\\medskip
J033219.77-274204.0&0.509$_{0.505}^{0.925}$&0.034$_{0.033}^{0.966}$&0.50&&\multicolumn{6}{c}{\textit{Failed  F$_{crit}$}}\\\medskip
J033220.02-274104.2&0.005$_{0.004}^{0.198}$&0.000$_{0.000}^{0.003}$&1.30&&9.06$\pm$0.130&2.70$\pm$0.053&0.89$\pm$0.003&-31.4$\pm$1.09&20.11$\pm$0.01&{\bf 0.612}\\\medskip
J033220.09-274106.7&0.001$_{0.000}^{0.056}$&0.000$_{0.000}^{0.000}$&2.64&&9.17$\pm$0.147&5.05$\pm$0.128&0.62$\pm$0.003&-66.0$\pm$0.33&20.16$\pm$0.01&0.686\\\medskip
J033220.67-274446.4&0.102$_{0.063}^{0.079}$&0.003$_{0.002}^{0.004}$&3.62&&3.91$\pm$0.026&2.33$\pm$0.011&0.60$\pm$0.001&-87.8$\pm$0.17&20.39$\pm$0.00&0.670\\\medskip
J033221.28-274435.6&0.509$_{0.105}^{0.132}$&0.022$_{0.010}^{0.024}$&1.76&&0.64$\pm$0.004&6.31$\pm$0.014&0.49$\pm$0.001&-35.1$\pm$0.12&20.36$\pm$0.00&{\bf 13.968}\\\medskip
J033222.33-274226.5&5.000$_{4.999}^{1.000}$&0.030$_{0.030}^{0.970}$&1.14&&3.63$\pm$0.050&2.02$\pm$0.013&0.43$\pm$0.003&60.8$\pm$0.20&21.28$\pm$0.00&0.421\\\medskip
J033222.58-274141.2&0.571$_{0.250}^{1.129}$&0.036$_{0.022}^{0.444}$&0.20&&2.15$\pm$0.032&2.53$\pm$0.013&0.88$\pm$0.003&-85.5$\pm$0.93&21.49$\pm$0.00&{\bf 0.666}\\\medskip
J033222.58-274152.1&0.255$_{0.052}^{0.031}$&0.740$_{0.520}^{0.260}$&0.84&&\multicolumn{6}{c}{\textit{Failed  F$_{crit}$}}\\\medskip
J033223.01-274331.5&0.806$_{0.520}^{1.094}$&0.076$_{0.064}^{0.924}$&0.97&&6.13$\pm$0.084&2.28$\pm$0.031&0.84$\pm$0.004&-74.3$\pm$0.88&20.84$\pm$0.01&0.426\\\medskip
J033224.36-274315.2&0.019$_{0.014}^{0.003}$&1.000$_{0.988}^{0.000}$&9.69&&\multicolumn{6}{c}{\textit{Failed  F$_{crit}$}}\\\medskip
J033224.98-274101.5&0.102$_{0.044}^{0.042}$&0.006$_{0.003}^{0.003}$&1.80&&6.05$\pm$0.035&1.82$\pm$0.009&0.82$\pm$0.001&44.6$\pm$0.32&20.03$\pm$0.00&0.506\\\medskip
J033225.11-274425.6&0.031$_{0.026}^{0.016}$&0.009$_{0.009}^{0.012}$&2.11&&\multicolumn{6}{c}{\textit{Failed  F$_{crit}$}}\\\medskip
J033225.29-274224.2&0.404$_{0.044}^{0.049}$&0.920$_{0.480}^{0.080}$&2.71&&\multicolumn{6}{c}{\textit{Failed  F$_{crit}$}}\\\medskip
J033225.47-274327.6&0.404$_{0.202}^{0.167}$&0.009$_{0.007}^{0.014}$&3.52&&3.74$\pm$0.017&5.95$\pm$0.022&0.87$\pm$0.001&71.2$\pm$0.30&21.49$\pm$0.00&{\bf 0.538}\\\medskip
J033225.85-274246.1&0.014$_{0.013}^{0.041}$&0.000$_{0.000}^{0.001}$&1.40&&\multicolumn{6}{c}{\textit{Failed F$_{crit}$}}\\\medskip
J033225.97-274312.5&0.453$_{0.452}^{2.297}$&0.009$_{0.009}^{0.990}$&0.36&&2.40$\pm$0.131&0.82$\pm$0.011&0.57$\pm$0.011&-33.9$\pm$1.22&22.77$\pm$0.01&0.443\\\medskip
J033225.98-274318.9&0.006$_{0.005}^{0.058}$&0.000$_{0.000}^{0.000}$&1.48&&6.08$\pm$0.129&1.59$\pm$0.026&0.44$\pm$0.005&-45.2$\pm$0.34&21.46$\pm$0.01&0.518\\\medskip
J033226.05-274236.5&5.000$_{0.750}^{0.000}$&1.000$_{0.400}^{0.000}$&2.76&&2.31$\pm$0.093&2.02$\pm$0.029&0.24$\pm$0.002&-9.1$\pm$0.19&23.05$\pm$0.03&{\bf 0.445}\\\medskip
J033226.71-274340.2&0.321$_{0.118}^{0.132}$&0.012$_{0.006}^{0.012}$&0.36&&3.65$\pm$0.063&1.35$\pm$0.011&0.58$\pm$0.002&42.9$\pm$0.27&22.65$\pm$0.01&{\bf 0.467}\\\medskip
J033227.18-274416.5&0.571$_{0.062}^{0.000}$&0.120$_{0.054}^{0.040}$&3.69&&4.81$\pm$0.028&4.17$\pm$0.010&0.56$\pm$0.000&29.0$\pm$0.05&21.16$\pm$0.01&{\bf 1.034}\\\medskip
J033227.62-274144.9&0.203$_{0.042}^{0.052}$&0.018$_{0.006}^{0.014}$&2.75&&1.81$\pm$0.028&1.85$\pm$0.008&0.68$\pm$0.002&-44.8$\pm$0.28&21.66$\pm$0.01&{\bf 0.640}\\\medskip
J033227.70-274043.7&1.015$_{0.375}^{0.885}$&0.120$_{0.084}^{0.880}$&1.34&&4.03$\pm$0.078&2.46$\pm$0.020&0.33$\pm$0.001&-77.0$\pm$0.10&22.17$\pm$0.02&{\bf 0.478}\\\medskip
J033227.84-274136.8&0.050$_{0.047}^{0.153}$&0.001$_{0.000}^{0.007}$&1.83&&6.75$\pm$0.075&10.16$\pm$0.21&60.61$\pm$0.002&52.7$\pm$0.27&20.30$\pm$0.01&0.492\\\medskip
J033227.86-274313.6&0.019$_{0.016}^{0.062}$&0.000$_{0.000}^{0.005}$&1.43&&\multicolumn{6}{c}{\textit{Failed F$_{crit}$}}\\\medskip
J033228.88-274129.3&0.203$_{0.164}^{0.202}$&0.003$_{0.003}^{0.010}$&0.64&&2.75$\pm$0.017&3.28$\pm$0.007&0.91$\pm$0.001&-56.4$\pm$0.59&21.88$\pm$0.01&{\bf 0.542}\\\medskip
J033229.04-274432.2&5.000$_{4.999}^{0.000}$&1.000$_{1.000}^{0.000}$&0.58&&2.20$\pm$0.096&1.32$\pm$0.016&0.20$\pm$0.010&-51.3$\pm$0.42&22.34$\pm$0.00&0.922\\\medskip
J033229.30-274244.8&0.072$_{0.052}^{0.089}$&0.004$_{0.003}^{0.009}$&0.89&&1.25$\pm$0.058&2.16$\pm$0.022&0.55$\pm$0.005&82.7$\pm$0.58&22.45$\pm$0.01&{\bf 0.503}\\\medskip
J033229.64-274030.3&0.102$_{0.062}^{0.101}$&0.009$_{0.006}^{0.026}$&2.08&&1.32$\pm$0.036&2.02$\pm$0.013&0.43$\pm$0.005&87.3$\pm$0.40&22.39$\pm$0.00&0.449\\\medskip
J033230.56-274145.7&0.143$_{0.042}^{0.059}$&0.042$_{0.020}^{0.046}$&1.82&&0.95$\pm$0.012&1.63$\pm$0.005&0.58$\pm$0.002&78.4$\pm$0.26&21.64$\pm$0.00&0.361\\\medskip
J033231.84-274329.4&0.002$_{0.001}^{0.225}$&0.000$_{0.000}^{0.009}$&0.58&&7.66$\pm$0.272&3.38$\pm$0.164&0.90$\pm$0.011&-78.4$\pm$3.61&21.71$\pm$0.02&0.393\\\medskip
J033232.34-274345.8&0.114$_{0.050}^{0.029}$&0.040$_{0.024}^{0.056}$&1.58&&\multicolumn{6}{c}{\textit{Failed F$_{crit}$}}\\\medskip
J033232.57-274133.8&0.001$_{0.000}^{0.001}$&0.002$_{0.000}^{0.000}$&2.37&&\multicolumn{6}{c}{\textit{Not Fit}}\\\medskip
J033232.96-274106.8&0.143$_{0.016}^{0.037}$&0.048$_{0.014}^{0.026}$&0.57&&0.78$\pm$0.015&1.15$\pm$0.005&0.89$\pm$0.002&69.8$\pm$1.37&23.30$\pm$0.02&{\bf 0.480}\\\medskip
J033233.28-274236.0&5.000$_{4.999}^{0.000}$&0.180$_{0.180}^{0.820}$&0.10&&2.59$\pm$0.169&1.18$\pm$0.023&0.39$\pm$0.014&38.6$\pm$0.92&23.03$\pm$0.01&0.464\\\medskip
J033233.40-274138.9&0.203$_{0.089}^{0.084}$&0.016$_{0.010}^{0.026}$&2.20&&1.87$\pm$0.023&2.25$\pm$0.007&0.82$\pm$0.002&-21.0$\pm$0.48&23.37$\pm$0.03&{\bf 0.388}\\\medskip
J033233.87-274357.6&0.005$_{0.004}^{0.250}$&0.000$_{0.000}^{0.007}$&1.48&&3.62$\pm$0.070&1.13$\pm$0.008&0.91$\pm$0.005&53.1$\pm$2.57&21.53$\pm$0.00&0.458\\\medskip
J033234.34-274350.1&0.161$_{0.059}^{0.094}$&0.009$_{0.004}^{0.010}$&3.56&&9.73$\pm$0.086&5.43$\pm$0.093&0.84$\pm$0.002&28.7$\pm$0.51&19.77$\pm$0.01&0.430\\\medskip
J033235.10-274410.7&0.052$_{0.026}^{0.028}$&0.012$_{0.006}^{0.010}$&8.73&&4.86$\pm$0.128&3.17$\pm$0.094&0.83$\pm$0.009&35.9$\pm$1.91&21.44$\pm$0.01&0.392\\\medskip
J033235.63-274310.2&0.001$_{0.000}^{0.025}$&0.000$_{0.000}^{0.000}$&1.16&&4.53$\pm$0.095&4.08$\pm$0.037&0.54$\pm$0.002&-24.8$\pm$0.22&22.43$\pm$0.01&{\bf 0.439}\\\medskip
J033236.72-274406.4&0.143$_{0.063}^{0.059}$&0.009$_{0.005}^{0.006}$&0.67&&3.79$\pm$0.086&4.87$\pm$0.065&0.54$\pm$0.003&23.9$\pm$0.28&22.96$\pm$0.02&{\bf 0.403}\\\medskip
J033237.32-274334.3&0.114$_{0.079}^{0.113}$&0.004$_{0.003}^{0.005}$&1.15&&1.73$\pm$0.020&4.37$\pm$0.043&0.87$\pm$0.005&-86.8$\pm$1.58&21.34$\pm$0.01&{\bf 0.493}\\\medskip
J033237.38-274126.2&0.404$_{0.149}^{0.167}$&0.007$_{0.005}^{0.012}$&1.72&&\multicolumn{6}{c}{\textit{Fail To Converge}}\\\medskip
J033238.06-274128.4&0.128$_{0.127}^{0.232}$&0.001$_{0.001}^{0.006}$&2.32&&5.80$\pm$0.032&4.76$\pm$0.037&0.58$\pm$0.001&40.0$\pm$0.12&19.67$\pm$0.00&0.535\\\medskip
J033238.36-274128.4&0.286$_{0.125}^{0.223}$&0.050$_{0.034}^{0.290}$&6.48&&3.35$\pm$0.060&1.21$\pm$0.008&0.90$\pm$0.004&46.2$\pm$2.49&24.37$\pm$0.05&{\bf 0.475}\\\medskip
J033238.44-274019.6&0.026$_{0.025}^{0.117}$&0.000$_{0.000}^{0.002}$&1.14&&6.01$\pm$0.074&3.09$\pm$0.042&0.72$\pm$0.003&-80.6$\pm$0.45&20.71$\pm$0.01&0.442\\\medskip
J033238.48-274313.8&0.255$_{0.052}^{0.066}$&0.160$_{0.078}^{0.360}$&1.66&&\multicolumn{6}{c}{\textit{Failed F$_{crit}$}}\\\medskip
J033239.17-274026.5&0.321$_{0.066}^{0.083}$&0.032$_{0.014}^{0.030}$&1.02&&2.41$\pm$0.026&3.83$\pm$0.014&0.65$\pm$0.001&-28.7$\pm$0.21&22.85$\pm$0.01&{\bf 0.448}\\\medskip
J033239.17-274257.7&0.571$_{0.211}^{0.236}$&0.032$_{0.018}^{0.040}$&0.54&&5.59$\pm$0.022&5.30$\pm$0.019&0.89$\pm$0.000&20.8$\pm$0.24&20.47$\pm$0.00&{\bf 0.562}\\\medskip
J033239.18-274329.0&0.003$_{0.002}^{0.224}$&0.000$_{0.000}^{0.006}$&0.44&&5.85$\pm$0.149&2.55$\pm$0.065&0.87$\pm$0.008&25.1$\pm$2.21&21.81$\pm$0.01&0.360\\\medskip
J033239.52-274117.4&0.052$_{0.052}^{0.203}$&0.000$_{0.000}^{0.005}$&1.42&&1.41$\pm$0.022&3.57$\pm$0.019&0.63$\pm$0.003&4.7$\pm$0.33&22.26$\pm$0.01&{\bf 0.457}\\\medskip
J033240.38-274338.3&0.025$_{0.022}^{0.047}$&0.000$_{0.000}^{0.001}$&0.65&&4.30$\pm$0.048&5.68$\pm$0.083&0.64$\pm$0.003&26.2$\pm$0.37&20.90$\pm$0.01&0.376\\\medskip
J033241.63-274151.5&0.038$_{0.034}^{0.052}$&0.000$_{0.000}^{0.001}$&4.30&&1.99$\pm$0.036&3.70$\pm$0.020&0.94$\pm$0.004&9.3$\pm$2.85&22.70$\pm$0.01&{\bf 0.474}\\\medskip
J033242.36-274238.0&0.509$_{0.149}^{0.132}$&0.018$_{0.010}^{0.018}$&1.51&&8.65$\pm$0.048&5.77$\pm$0.056&0.46$\pm$0.000&74.4$\pm$0.06&18.79$\pm$0.00&0.977\\\medskip
J033243.93-274232.4&0.001$_{0.000}^{0.063}$&0.000$_{0.000}^{0.000}$&3.03&&3.54$\pm$0.056&4.84$\pm$0.039&0.59$\pm$0.001&28.6$\pm$0.25&22.49$\pm$0.01&{\bf 0.478}\\\medskip
J033244.97-274309.1&0.009$_{0.008}^{0.171}$&0.000$_{0.000}^{0.002}$&4.93&&\multicolumn{6}{c}{\textit{Failed F$_{crit}$}}\medskip
\label{tab:yspfits} 
\end{longtable} 
\endgroup 
 
\clearpage \newpage 

\clearpage
\newpage
\begingroup
\tiny 
\begin{longtable}{cccc}
\caption{Number of Likely Companions}\\
\hline \hline 
\multicolumn{1}{c}{GOODS ID} &
\multicolumn{1}{c}{$N_c$} &
\multicolumn{1}{c}{$n_{spec}$} &
\multicolumn{1}{c}{$n_{phot}$} \\ \hline \hline
\endfirsthead
\\ \\
\multicolumn{4}{c}{Table 2: Companion Number, cont.}\\ 
\hline \hline 
\multicolumn{1}{c}{GOODS ID} &
\multicolumn{1}{c}{$N_c$} &
\multicolumn{1}{c}{$n_{spec}$} &
\multicolumn{1}{c}{$n_{phot}$} \\ \hline \hline
\endhead

\\ \hline
\multicolumn{4}{l}{Continued on next page ...}\\
\endfoot
\hline \hline
\multicolumn{4}{l}{\textbf{Notes-} Column 2$:$ Number of likely companions,}\\
\multicolumn{4}{l}{~~~~~with 1$\sigma$ uncertainties measured from an}\\
\multicolumn{4}{l}{~~~~~empirical jackknife technique(\S\ref{subsec:compnomeas}).}\\
\multicolumn{4}{l}{~~~~~Cols. 3 and 4$:$ number of photometric}\\
\multicolumn{4}{l}{~~~~~and spectroscopic possible companions.}
\endlastfoot
J033205.09-274514.0&3.03$\pm$1.64&7&3\\ \smallskip  
J033206.27-274536.7&1.01$\pm$0.82&2&1\\ \smallskip  
J033207.55-274356.6&1.02$\pm$0.90&4&1\\ \smallskip  
J033208.53-274217.7&1.01$\pm$0.87&3&1\\ \smallskip  
J033210.04-274333.1&1.01$\pm$0.77&3&1\\ \smallskip  
J033210.12-274333.3&1.02$\pm$0.91&5&1\\ \smallskip  
J033211.21-274533.4&1.01$\pm$0.87&3&1\\ \smallskip  
J033211.61-274554.1&1.01$\pm$0.82&2&1\\ \smallskip  
J033212.20-274530.1&1.01$\pm$0.82&2&1\\ \smallskip  
J033212.31-274527.4&1.02$\pm$0.87&3&1\\ \smallskip  
J033214.45-274456.6&1.01$\pm$0.89&4&1\\ \smallskip  
J033219.02-274242.7&1.01$\pm$0.87&3&1\\ \smallskip  
J033226.71-274340.2&1.01$\pm$0.93&6&1\\ \smallskip  
J033231.84-274329.4&1.02$\pm$0.90&4&1\\ \smallskip  
J033233.40-274138.9&2.01$\pm$1.22&2&2\\ \smallskip  
J033235.10-274410.7&1.00$\pm$0.82&2&1\smallskip     
\label{tab:tablex} 
\end{longtable} 
\endgroup 

  
\begin{figure}[htbc]
\begin{center}
\epsfig{file=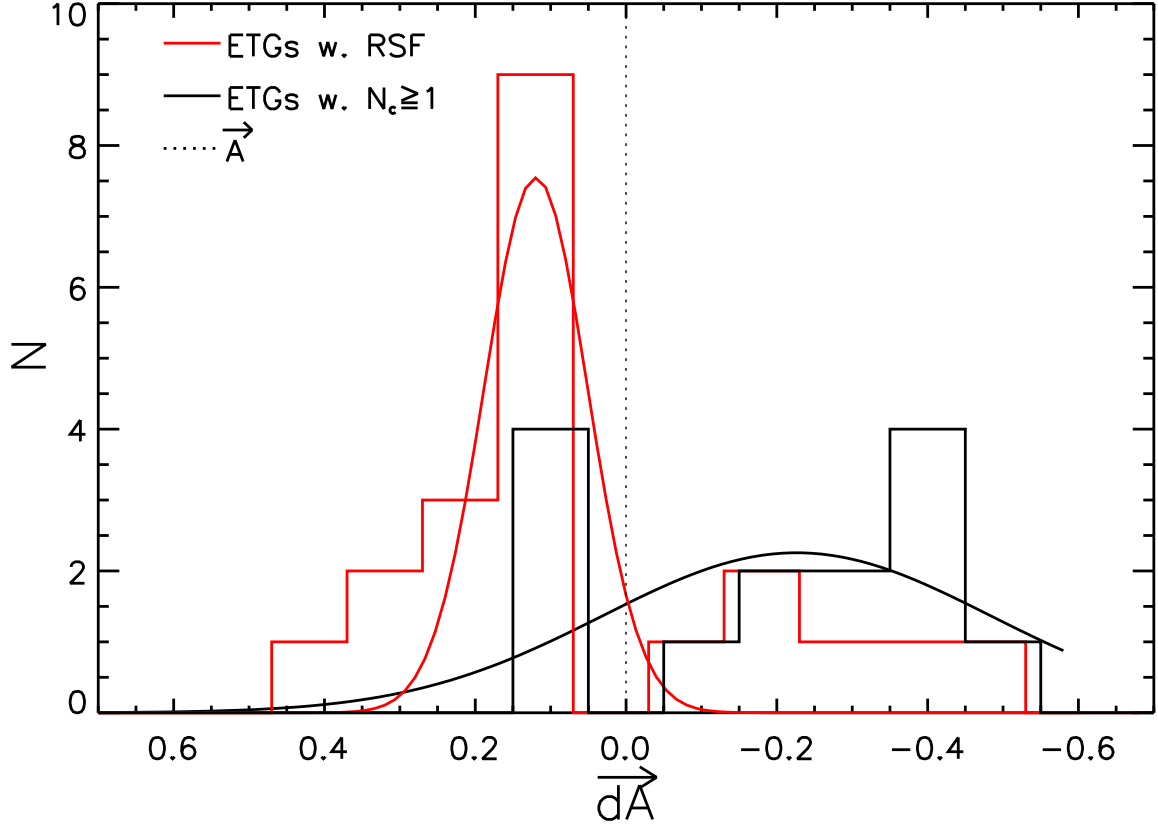,width=1.0\textwidth,clip=}

\caption{In \S\ref{subsec:discussion3a}, we noted that high mass
(M\gsim10.5\Msol) ETGs with recent star-formation appear to be
distributed near the low-redshift ($z\sim0$) empirical size-mass
relationship. In contrast, quiescent ETGs and particularly ETGs with
$N_c\ge1$ cluster, though with larger dispersion, near the intermediate
redshift ($1.0<z<1.5$) empirical size-mass relationship. Here, we plot
histograms of these populations' perpendicular distances, $\vec{dA}$,
from the preferred vector, $\vec{A}$ (see Fig. 5) for ETGs with
$N_c\ge1$ (black) and RSF (red), respectively, with Gaussian fits to
each distribution overplotted.  In \S\ref{subsec:discussion3b}, we find
these means of each distributions are distinguishable by a two-sample
t-test, i.e., the null hypothesis is rejected at \gsim95\%.  We discuss
the implications of this distinction in \S\ref{subsec:discussion3b}, with respect to the mechanism for
inducing RSF in intermediate-redshift ETGs.} \label{fig:SFQdistinct}
\end{center} \end{figure}


\end{document}